\documentclass[10pt,journal,compsoc]{IEEEtran}

\ifCLASSOPTIONcompsoc
\usepackage[nocompress]{cite}
\else
\usepackage{cite}
\fi
\usepackage[breaklinks]{hyperref}
\usepackage{microtype}
\usepackage{hyperref} 
\usepackage{graphicx}
\usepackage{amssymb}
\usepackage{color}
\usepackage{epstopdf}
\usepackage{array}
\usepackage{multirow}
\usepackage{amsfonts}
\usepackage{booktabs}
\usepackage{float}
\usepackage[caption=false]{subfig}
\usepackage{algorithm}
\usepackage{algpseudocode}
\usepackage[switch]{lineno}
\usepackage{comment}
\usepackage[utf8]{inputenc}
\usepackage[T1]{fontenc}
\usepackage{enumitem}
\usepackage{listings}  
\usepackage{graphics}
\usepackage{graphicx}
\usepackage{csquotes}
\usepackage{longtable}
\usepackage[fleqn]{amsmath}
\usepackage{siunitx}
\usepackage{nicefrac} 
\usepackage{fancyvrb}
\usepackage{lipsum}
\usepackage{tabularx,multirow}
\begin{document}
%
%
\title{Deep ROC Analysis and AUC as Balanced Average Accuracy to Improve Model Selection, Understanding and Interpretation}
%
\author{Andr\'e M. Carrington, Douglas G. Manuel, Paul W. Fieguth, Tim Ramsay, Venet Osmani, \\ Bernhard Wernly, Carol Bennett, Steven Hawken, Matthew McInnes, Olivia Magwood, Yusuf Sheikh, \\ and Andreas Holzinger,~\IEEEmembership{Member,~IEEE}
\IEEEcompsocitemizethanks
{\IEEEcompsocthanksitem 
A.M. Carrington is with the Ottawa Hospital Research Institute and the Institute of Clinical Evaluative Sciences, Ottawa, Canada. E-mail: acarrington@ohri.ca
\IEEEcompsocthanksitem 
D.G. Manuel is with the Ottawa Hospital Research Institute; the Department of Family Medicine and the School of Epidemiology and Public Health, University of Ottawa; the Institute of Clinical Evaluative Sciences; the Bruyère Research Institute.  E-mail: dmanuel@ohri.ca
\IEEEcompsocthanksitem 
P.\ Fieguth is with the Department of Systems Design Engineering, co-directs the Vision and Image Processing Lab, and is Associate Dean in the Faculty of Engineering, University of Waterloo, Canada. E-mail: paul.fieguth@uwaterloo.ca
\IEEEcompsocthanksitem 
V. Osmani is with the e-health group at Fondazione Bruno Kessler Research Institute and the department of Psychology and Cognitive Science at University of Trento, Italy. E-mail: vosmani@fbk.eu
\IEEEcompsocthanksitem 
B. Wernly is with the Department of Cardiology, Paracelsus Medical University of Salzburg, Salzburg, Austria. E-mail: b.wernly@salk.at
\IEEEcompsocthanksitem 
S. Hawken is with the Ottawa Hospital Research Institute and the University of Ottawa. E-mail: shawken@ohri.ca
\IEEEcompsocthanksitem 
M. McInnes is with the Ottawa Hospital Research Institute and the University of Ottawa. E-mail: mmcinnes@toh.on.ca
\IEEEcompsocthanksitem 
T. Ramsay is with the Ottawa Hospital Research Institute and the University of Ottawa. E-mail: tramsay@ohri.ca
\IEEEcompsocthanksitem 
C. Bennett is with the Ottawa Hospital Research Institute and the Institute of Clinical Evaluative Sciences, Ottawa, Canada. E-mail: cbennett@ohri.ca
\IEEEcompsocthanksitem 
O. Magwood is with the Bruyere Research Institute and is a doctoral student at the University of Ottawa, Canada. E-mail: omagwood@bruyere.org
\IEEEcompsocthanksitem 
Y. Sheikh is with the Ottawa Hospital Research Institute and is an undergraduate student in the Department of Biology, University of Ottawa, Canada. E-mail: ysheikh@ohri.ca
\IEEEcompsocthanksitem 
A. Holzinger is with the Alberta Machine Intelligence Institute, University of Alberta, Canada and head of the Human-Centered AI Lab, Medical University Graz, Austria. E-mail: andreas.holzinger@medunigraz.at. 
\protect\\
\IEEEcompsocthanksitem Corresponding author: Andreas Holzinger.
}
}
%
\markboth{IEEE Transactions on Pattern Analysis and Machine Intelligence}
{Carrington et al.\MakeLowercase {
}: Deep ROC Analysis and Balanced Average Accuracy}
%
\IEEEtitleabstractindextext{
\begin{abstract}
Optimal performance is critical for decision-making tasks from medicine to autonomous driving, however common performance measures may be too general or too specific. For binary classifiers, diagnostic tests or prognosis at a timepoint, measures such as the area under the receiver operating characteristic curve, or the area under the precision recall curve, are too general because they include unrealistic decision thresholds. On the other hand, measures such as accuracy, sensitivity or the F1 score are measures at a single threshold that reflect an individual single probability or predicted risk, rather than a range of individuals or risk.  We propose a method in between, deep ROC analysis, that examines groups of probabilities or predicted risks for more insightful analysis. We translate esoteric measures into familiar terms: AUC and the normalized concordant partial AUC are balanced average accuracy (a new finding); the normalized partial AUC is average sensitivity; and the normalized horizontal partial AUC is average specificity.  Along with post-test measures, we provide a method that can improve model selection in some cases and provide interpretation and assurance for patients in each risk group. We demonstrate deep ROC analysis in two case studies and provide a toolkit in Python. 
\end{abstract}
}
\maketitle
%
\IEEEpeerreviewmaketitle

\IEEEraisesectionheading{\section{Introduction}
\label{sec:introduction}}
\IEEEPARstart{T}wo important and common measures of performance for binary diagnostic tests and classifiers (models) are accuracy \cite{Santafe2015} and the area under the curve \cite{Steyerberg2009b} (AUC) in a receiver operating characteristic (ROC) plot \cite{Fawcett2006}. Accuracy is a measure at a single operating point or decision threshold on a model's ROC curve, while AUC measures all operating points. 

In the medical and health domain a lot of data is imbalanced \cite{DubeyYe:2014:ImbalancedExample}. For imbalanced data, alternative measures 
at a single operating point include balanced accuracy \cite{Santafe2015}, the geometric mean (of sensitivity and specificity) \cite{Santafe2015,wu2009small}, the $F_1$ score \cite{Santafe2015,Flach2019} and Matthews' Correlation Coefficient \cite{Zhu2020}. At all operating points, a common alternative is the area under the precision recall curve (AUPRC) a.k.a. average precision (AP) \cite{Saito2015}; while less common alternatives include the predictive ROC curve \cite{shiu2008predictive}, the positive tradeoff (PT) curve \cite{OReilly2013}, the H measure \cite{Hand2009} and the area under the cost curve \cite{flach2011coherent}.

However, all measures of performance at a single operating point are too specific---they depend on a specific choice of misclassification costs that reflect a single or average patient, and lack information about performance at points nearby where performance may change rapidly \cite{McClish2012,Bradley1997}. 

On the other hand, all measures of performance at all operating points, i.e., global measures, are too general. AUC, a global measure, is preferred over accuracy \cite{Bradley1997} but AUC is criticized because it includes operating points that would not be used in practice \cite{Lobo2008,mallett2012interpreting} and it doesn't provide any information about the distribution of performance along the ROC curve \cite{Lobo2008}. 

ROC plots are intended to show the distribution of performance for further analysis \cite{Flach2019, Provost1998} but they are visual and do not provide a number of useful quantitative measures---e.g., what is the average sensitivity, AUC and positive predictive value within a group? Precision recall curve (PRC) plots 
have similar shortcomings.

ROC analysis is typically used to observe the dominance or rank of classifiers overall, to observe where dominance changes when ROC curves cross, or to choose an optimal ROC point or threshold \cite{Provost1998,Fawcett2006}.

\begin{table}[btp]
\captionsetup{justification=justified}
\caption{\label{tab:Example-table1}Consider a binary classifier or diagnostic test for data with $30\%$ prevalence. Suppose the high risk group is most relevant. AUC, as a global measure, obscures all of the group-wise measures.  Relative to the AUC, the high risk group has a better balanced average accuracy of $85\%$, but a significantly lower average sensitivity of $67\%$. The high risk group has the highest balanced average accuracy among groups---so the result may not improve by optimizing with different hyperparameters. Confidence intervals are omitted for simplicity and post-test measures are discussed in case studies (Sections \ref{sec:case1}, \ref{sec:case2}).}
\begin{tabular}{l|c|c|c|c}
\hline
ROC horizontal axis (FPR):       & Global  & Left  & Mid  & Right \\
                                 & [0,1]  & [0,.33] & [.33,.67] & [.67,1]\\
Probability/risk group:          & All    & High  & Med  & Low\\
\hline
Bal Avg Accuracy\,\,\, = $AUC$            & 0.82   &       &         &\\
Group Bal Avg Acc = $\widetilde{cpAUC}$ & 0.82   & 0.85  & 0.81    & 0.76\\
Group Avg Sens \quad = $\widetilde{pAUC}$& 0.82   & 0.67  & 0.84    & 0.94\\
Group Avg Spec \quad = $\widetilde{pAUCx}$& 0.82 & 0.93  & 0.67    & 0.40\\
\hline
Positive predictive value & \multicolumn{4}{c}{\multirow{2}{*}{0.48 (at threshold=0.5)}}\\
at a point & \multicolumn{4}{c}{}\\
\hline
\end{tabular}
\end{table}

We posit the need for \textbf{deep ROC analysis}---a quantitative analysis of ROC data based on explicitly specified groups of probability or risk (Table \ref{tab:Example-table1}). In comparison to global measures of performance, or performance at a point, deep ROC analysis can lead to different decisions to select or accept a binary classifier or test.

In our proposed method one may use as many risk groups as needed, only limited by the number of instances (e.g., patients) in the data. The risk groups may be percentiles of predicted risk or probability, intervals in specificity (or its complement, $\textit{FPR}$), or intervals in sensitivity ($\textit{TPR}$).

Support for our group-wise approach can be found in a recent systematic review. Wynants \textit{et al.} \cite{wynants2020prediction} examined over 100 COVID-19 prediction models and recommended that none of the models be used in practice, in part because of lack of reporting on calibration.  Calibration measures performance by groups, similar to our proposed deep ROC analysis, except our method focuses on measures of discrimination, as distinct from calibration \cite{Steyerberg2009a}.

Two key contributions of deep ROC analysis are:
\begin{enumerate}
\item \textbf{properly measuring AUC in groups with the normalized concordant partial AUC ($\mathbf{\widetilde{cpAUC}}$) \cite{CarringtonEtAl:2020:AUC}}
\item \textbf{a new interpretation of AUC and $\mathbf{\widetilde{cpAUC}}$ as balanced average accuracy}
\end{enumerate}

To compare groups organized left to right in an ROC plot (Table \ref{tab:Example-table1}) we cannot use the group averages for sensitivity which always increase to the right nor the group averages for specificity which always increase to the left. We need the concept of AUC within a group which is fulfilled by $\widetilde{cpAUC}$. In the Related Work and Background sections that follow we explain why alternatives are improper or insufficient.

Also, current interpretations of AUC are lacking and abstract \cite{mallett2012interpreting, wagstaff2012machine}. If you ask someone what does an AUC of $0.8$ or $80\%$ mean? Or what does a $2\%$ improvement in AUC mean? The two most common answers are as follows. 

First, one might receive a comparative explanation: that an AUC of 0.5 indicates a classifier (or test) is no better than chance, whereas an AUC of 1.0 means the classifier is perfect at discrimination.  As the name indicates AUC is the area under the ROC curve which is depicted in an ROC plot. Considering the plot's axes of sensitivity and 1-specificity, the AUC represents how sensitive and specific a classifier or test is at many different operating points along the ROC curve. However this explanation does not tell us what an AUC of $0.8$ means precisely: how many errors will the classifier or test commit and in which subgroups?

The second more precise answer is that the AUC can be interpreted as a C statistic: the likelihood that the classifier ranks (scores) a randomly chosen positive patient higher than a randomly chosen negative patient. Therefore, an AUC of $80\%$ means that the classifier is correct $80\%$ of the time in pairwise ranking; and a $2\%$ improvement means that in pairwise ranking, the classifier is correct $2\%$ more often---which seems meaningful at first, but ranking is not decision-making.  What is the probability of error for a single patient? What is the probability of error for a subgroup of patients (e.g., those who are predicted with high probability of having the condition)?

Consider, if the $2\%$ improvement only ranks low-risk patients better against each other, or only ranks high-risk patients better against each other, then that improvement may not change our decision-making that distinguishes high from low risk, nor the classifier’s output. A classifier is concerned with discriminating those with a condition from those without. 
Hence, we seek a better interpretation than the C statistic's concept of pairwise-ranking. 

Two other interpretations of the AUC are that: AUC equals average sensitivity across all thresholds, and AUC equals average specificity across all thresholds \cite{ZhouEtAl:2002:StatisticalBook}---as observed in the Global column of Table \ref{tab:Example-table1}.

Hence a classifier with an AUC that is $2\%$ higher, is on average, over all possible thresholds, $2\%$ more sensitive at detecting positives and $2\%$ more specific (i.e., it detects negatives $2\%$ better). These equalities are not true for part of an ROC curve, however, where average sensitivity and average specificity, in general, differ \cite{CarringtonEtAl:2020:AUC}.

What does hold true, is that the average (or balance) of average sensitivity and average specificity, i.e., \textbf{balanced average accuracy, is equal to AUC} (Section \ref{sec:AUC-as-BAA}); and for part of an ROC curve, \textbf{balanced average accuracy is equal to $\mathbf{\widetilde{cpAUC}}$} (Section \ref{sec:cpAUCn-as-BAA}).

Our finding on balanced average accuracy is not to be confused with a previously-known special case. When an ROC curve consists of a single point, $S$, aside from the peripheral endpoints $(0, 0)$ and $(1, 1)$, then $\textrm{AUC}_S$ in that special case is equal to balanced accuracy (\ref{eq:balanced_accuracy_w}) at the point $S$ \cite{sokolova2006beyond}. This special case occurs for discrete classifiers \cite{Fawcett2006}, e.g., a decision rule or decision tree. 

In the sections that follow we discuss related work, background, our method, two case studies, limitations, conclusions and future work.

\section{Related Work}
\label{sec:relatedwork}
Our work seeks to understand and interpret model performance with AUC and related measures in greater detail with a new method and a new interpretation of AUC.

ROC analysis has become a standard tool in the design and evaluation of two-class classification problems \cite{obuchowski2018receiver} with ongoing work \cite{xue2014does, Landgrebe:2008:MulticlassROC} and extensions \cite{perez2018nsroc}. This is because it allows us to analyze operating points and incorporate costs and priors—which are important issues for many real-world problems where conditions are often non-ideal (non i.i.d.).  Analysis of models with ROC plots is a topic with continued growth in the statistical literature \cite{perez2018nsroc}.

Great advances have been made in machine learning and particularly in deep learning applied to various fields of medicine and smart health with high accuracy \cite{XingLing:2017:DeepMedExample}, \cite{ZuoLing:2017:Success}. To make such successes even more successful the field of explainable artificial intelligence (xAI) is attracting much interest in the health domain \cite{HolzingerEtAl:2019:Wiley-Paper}, \cite{TjoaErico:2021:MedicalxAI}.

The xAI community is supporting such efforts in developing methods that provide transparency and traceability for such deep learning approaches which are considered as statistical "black-box" methods. \cite{HolzingerKieseWeipplTjoa:2018:trends}. 
Recent work on a large-scale nonlinear AUC maximization method (called TSAM) based on triply stochastic gradient descents is relevant for ROC and performance analysis in machine learning generally and for explainable AI specifically \cite{DangDeng:2020:AUCnew}.

There is literature advising on problems to avoid with measures \cite{Flach2019, vickers2010everything, mallett2012interpreting} and there are surveys of available measures \cite{Santafe2015, OReilly2013, Powers2007}. However, there is less literature on how to best use measures together, i.e., overall methods, for greater insight and effectiveness.

\begin{figure*}[btp]
\captionsetup[subfigure]{labelformat=brace}
\subfloat[The partial AUC ($pAUC$) is a vertical slice of the area under the ROC curve (AUC) and its normalized value ($\widetilde{pAUC}$) is the average sensitivity (height)]{
\centering
    \includegraphics[scale=0.35]{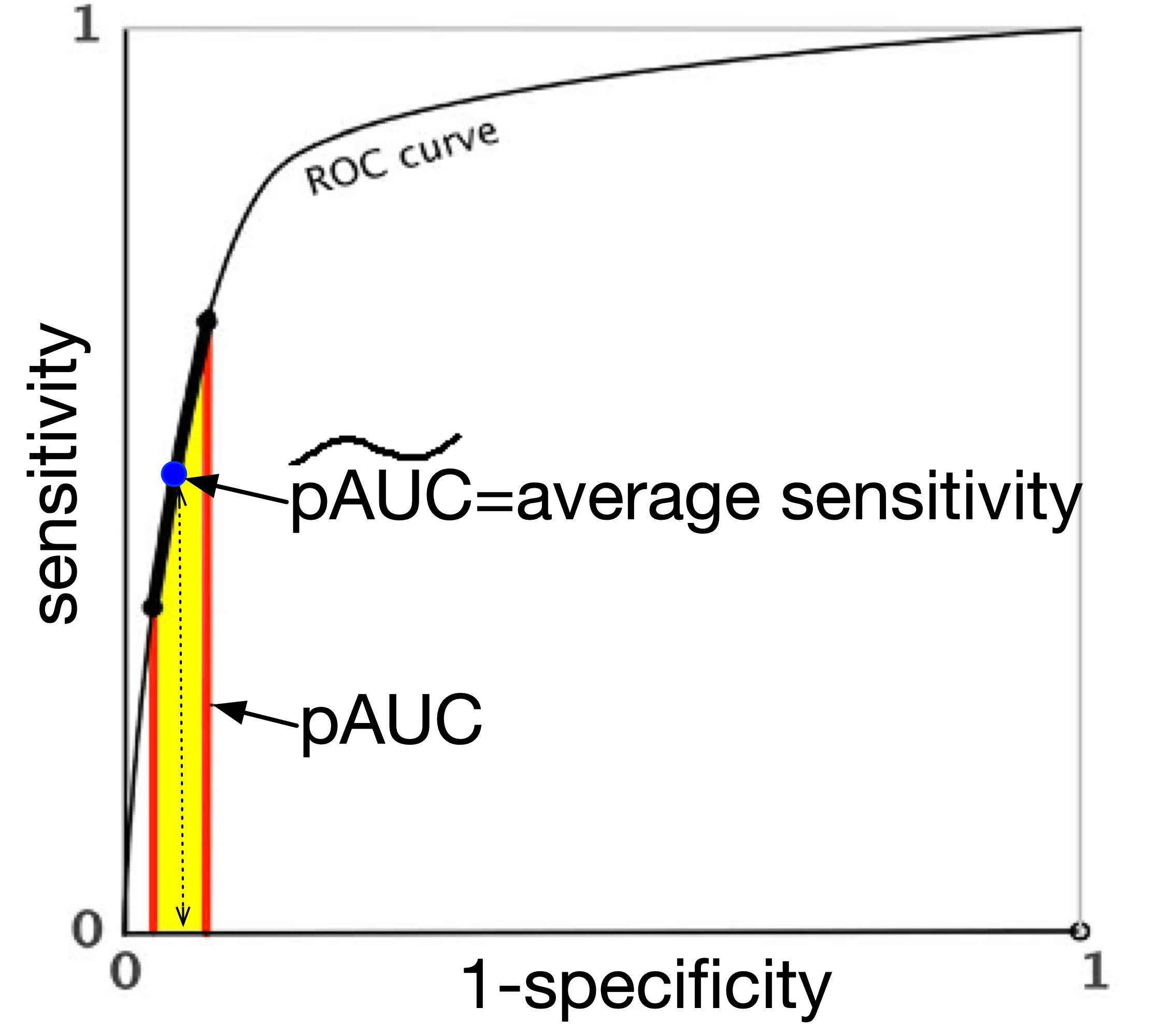}
}
\hskip6.3em
\captionsetup[subfigure]{labelformat=brace}
\subfloat[The horizontal partial AUC ($pAUCx$) is a horizontal slice of the AUC and its normalized value ($\widetilde{pAUCx}$) is the average specificity (width)]{
\centering
    \includegraphics[scale=0.35]{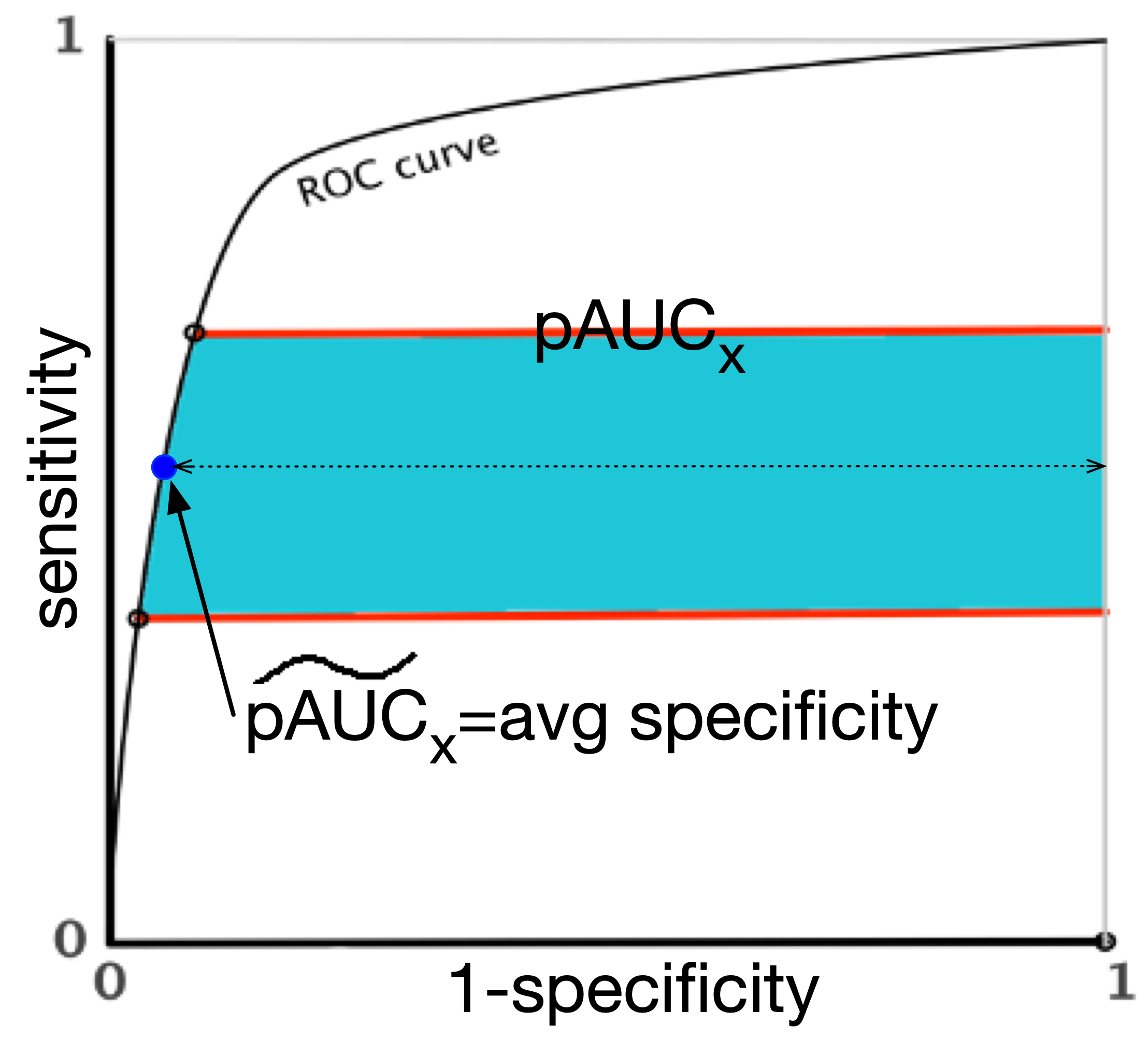}
}
\centering
\captionsetup{margin=10pt}
\caption{Two measures used in our method represent average sensitivity and average specificity, but are more commonly known by esoteric labels. Analysis is made complete by balanced average accuracy, as a third measure. \label{fig:pAUC_pAUCx}}
\end{figure*}

Sokolova \textit{et al.} \cite{sokolova2006beyond} argue that performance measures commonly used in machine learning do not properly address situations where the classes are equally important and several models are compared. They propose three measures: Youden's index\footnote{Youden's index is linearly related to balanced accuracy at the point where the ROC curve intersects the minor diagonal.}, likelihood ratios, and discriminant power---but these measures are not popular. %
Sokolova and Lapalme \cite{sokolova2009systematic} survey the invariant properties of performance measures and recommend measures for natural language processing.

Steyerberg \textit{et al.} \cite{Steyerberg2009b} recommend reporting measures of discrimination and calibration, including the C statistic or AUC, and measures from a calibration plot. These measures are popular, and our method supplements them with deeper analysis. For clinical decision-making they also recommend net benefit as a measure of clinical utility. Steyerberg and Vergouwe \cite{Steyerberg2014} re-iterate the same categories but narrow down the measures a little further from six to four\footnote{Although a fifth measure, the odds ratio, is also discussed.}. 

Mallett \textit{et al.} \cite{mallett2012interpreting} discuss various measures of discrimination and clinical utility, including the partial AUC's benefit over AUC, but they do not discuss measuring multiple groups and they do not provide an overall method. Obuchowski and Bullen \cite{obuchowski2018receiver} provide a survey of case studies or applications of AUC and related measures but they do not provide a general method to follow.  

Several reporting guidelines have been produced to improve the completeness and transparency of published studies of diagnostic tests and prediction models. For example, STARD asks authors to report their positivity cut-offs, how they were determined and whether they were defined a priori \cite{cohen2016stard}. Similarly, TRIPOD asks authors to define all predictors and the outcome that is predicted by the prediction model, including how and when they were measured \cite{collins2015transparent}. Across all reporting guidelines published to date, none accommodate personalized medicine with classifiers whose thresholds can be tuned or re-calibrated at the point of service specific to the setting, a group, or even a patient. How certain can we be that a test will be suitable for a given population? A GRADE assessment represents our confidence that the true accuracy of a diagnostic test lies above or below a threshold, or in a specified range \cite{hultcrantz2020defining} that depends on prevalence. The range may also consider the cost of a test's direct effects and its downstream health consequences of true and false positives/negatives \cite{hultcrantz2020defining}.

Since our proposed method examines performance by groups of risk, or parts of the ROC and AUC data, we discuss precedents for that approach. 

Examples of ROC analysis in the literature that applied a group-wise approach include Provost \textit{et al.} who describe the dominant classifier in groups by slope (or skew) where a different classifier dominates in each group \cite{Provost1998}. Dominance ensures better performance by a variety of common measures: accuracy, sensitivity, specificity, balanced accuracy, positive predictive value, etc. However, the question arises: how much better is the performance? Provost \textit{et al.} do not quantify the difference, but they show confidence intervals toward that interest.

Bradley \cite{bradley2014half} provides an alternative, the 
half-AUC, to examine the area in an ROC plot in two parts, separated by the minor diagonal which extends from the top left to the bottom right, and where sensitivity and specificity are separately emphasized in each part. This approach is sensible, but limits analysis to two groups with fixed bounds. Also, while it is scaled to the same range as the AUC or C statistic, it is not shown to have the same or comparable meaning.

Other examples are Carrington \textit{et al.} \cite{CarringtonEtAl:2020:AUC} and Wernly \textit{et al.} \cite{wernly2020machine} who compare classifiers by the partial AUC and the concordant partial AUC in groups by false positive rate---but these measures are not popular or familiar. When those two measures are normalized, however, they are familiar as group average sensitivity and the group's AUC, respectively.

There is also ample literature on performance measures that may be used in a part or group-wise manner without attempting to provide a holistic method \cite{CarringtonEtAl:2020:AUC, yang2019two, bradley2014half, McClish2012, Wu2008, pepe2003statistical, Dodd2003, Jiang1996, McClish1989, Thomson1989}---we use and review some of these in the next section. 

On interpretations of the AUC, related work includes: the concordant partial AUC as a generalization of the AUC and its relation to the partial C statistic \cite{CarringtonEtAl:2020:AUC}, AUC related to utility
\cite{Hernandez-Orallo2012,flach2011coherent,Hand2009}, AUC related to AUPRC \cite{Su2015,Davis2006} and conceptual discussion of requirements for AUC related to utility \cite{shah2019making}.

Our new interpretations of AUC and our method also support causability for explainable medicine  \cite{HolzingerEtAl:2019:Wiley-Paper,HolzingerEtAl:2021:GraphFusion}, as a step beyond explainable AI \cite{carrington2018measures}. The term causability was coined in reference to the established term usability, and is defined as the measurable extent to which an explanation, from AI, considered by a human expert, achieves a specified level of causal understanding. The quality of explanation can be measured---e.g., with the System Causability Scale \cite{HolzingerEtAl:2020:QualityOfExplanations}---in the same way that usability encompasses measurements for the quality of use.

\section{Background}
\label{sec:background}

While Bradley \cite{Bradley1997} recommended AUC over accuracy, others have since identified issues with the area under the ROC curve (AUC) as a measure of performance \cite{McClish2012,wagstaff2012machine, Lobo2008}---and these criticisms also apply to the C statistic for binary outcomes\footnote{For empirical ROC curves and binary outcomes the AUC and C statistic are equal \cite{vickers2010everything,Steyerberg2009a,Cook2008}}. The C statistic for binary outcomes \cite{Pencina2015, Steyerberg2009b} we refer to should \textbf{not} be confused with Harrell or Uno's C statistics for continuous outcomes \cite{Guo2017,uno2011c,FE1982}.

Since AUC is an overall measure, McClish and, separately, Thomson and Zucchini \cite{McClish1989, Thomson1989} proposed the partial AUC ($pAUC$) (Figure \ref{fig:pAUC_pAUCx}a), which can be applied to any subset of the false positive rate (1 - specificity). This was a first step toward deep ROC analysis. A later definition of $pAUC$ used a non-parametric fit with fewer assumptions \cite{McClish1989}. When the $pAUC$ is normalized by its range for $\Delta x=x_2-x_1$ for an ROC curve $y=r(x)$ it becomes:
\begin{align}
     \widetilde{pAUC}(x_1,x_2) &= \frac{1}{\Delta x} \int_{x_1}^{x_2}r(x)\ \label{eq:pAUCn} dx
\end{align}
The name partial AUC is misleading because it does not have all of AUC's characteristics \cite{CarringtonEtAl:2020:AUC}.

Mallet \textit{et al.}, who promoted the use of $pAUC$ and compared two tests \cite[Fig. 3e,f]{mallett2012interpreting} with it, suggested "...the tests are equally effective" based on similar $pAUC$ values \cite[Pg. 4]{mallett2012interpreting}.  However, 
the tests had nearly identical sensitivity but starkly different specificity: $78-95\%$ versus $50-60\%$. Mallet \textit{et al.} provide no discussion or rationale for this discrepancy. $pAUC$ considers the width of the range of specificity but not its values.

Soon after, McClish \cite{McClish2012} acknowledged that $pAUC$ is flawed because it monotonically increases to the right in an ROC plot---and others also found fault with $pAUC$ \cite{ma2013use}. McClish \cite{McClish2012} therefore proposed the standardized Partial Area ($\mathit{sPA}$) which begins with the $pAUC$, subtracts the area under the major diagonal, and then standardizes the result. $\mathit{sPA}$ is intended for comparison to the AUC.

While $\mathit{sPA}$ eliminates or reduces monotonic behaviour, its approach is flawed \cite{CarringtonEtAl:2020:AUC,vivo2018rethinking} because it can produce a negative result for an ROC curve that is partly above the major diagonal and partly below it. Such ROC curves occur in real life \cite{vivo2018rethinking,perez2018nsroc,ZhouEtAl:2002:StatisticalBook,metz1980statistical}. Negative values of $\mathit{sPA}$ mean that $\mathit{sPA}$ cannot be interpreted as an AUC or C statistic, because the formulas for the latter are only additive; and other measures we will discuss are interpretable as an AUC or C statistic \cite{CarringtonEtAl:2020:AUC}.

If we return our attention to $pAUC$, it does not meet the requirements for an overall measure---but it is useful when properly applied with other measures. The partial AUC \cite{Dodd2003}, when normalized ($\widetilde{pAUC}$), is average sensitivity \cite{CarringtonEtAl:2020:AUC} and therefore has a vertical perspective (Figure \ref{fig:pAUC_pAUCx}a). 

The $pAUC$ has a horizontal counterpart: the partial area index ($PAI$) or the normalized horizontal partial AUC ($\widetilde{pAUCx}$) as average specificity  \cite{CarringtonEtAl:2020:AUC,Jiang1996} (Figure \ref{fig:pAUC_pAUCx}b) over the range $\Delta y = y_2-y_1$ for an ROC curve $x=r^{-1}(y)$:
\begin{align}
  \widetilde{pAUC_x}(y_1,y_2) &= \frac{1}{\Delta y} \int_{y_1}^{y_2}{1 - r^{-1}(y)}dy \label{eq:pAUCxn}
\end{align}
We apply both of these measures in our method (Table \ref{tab:Example-table1}) as well as the next measure.

\begin{figure}[btp]
\centering
    \includegraphics[width=2.6in,height=2.5in]{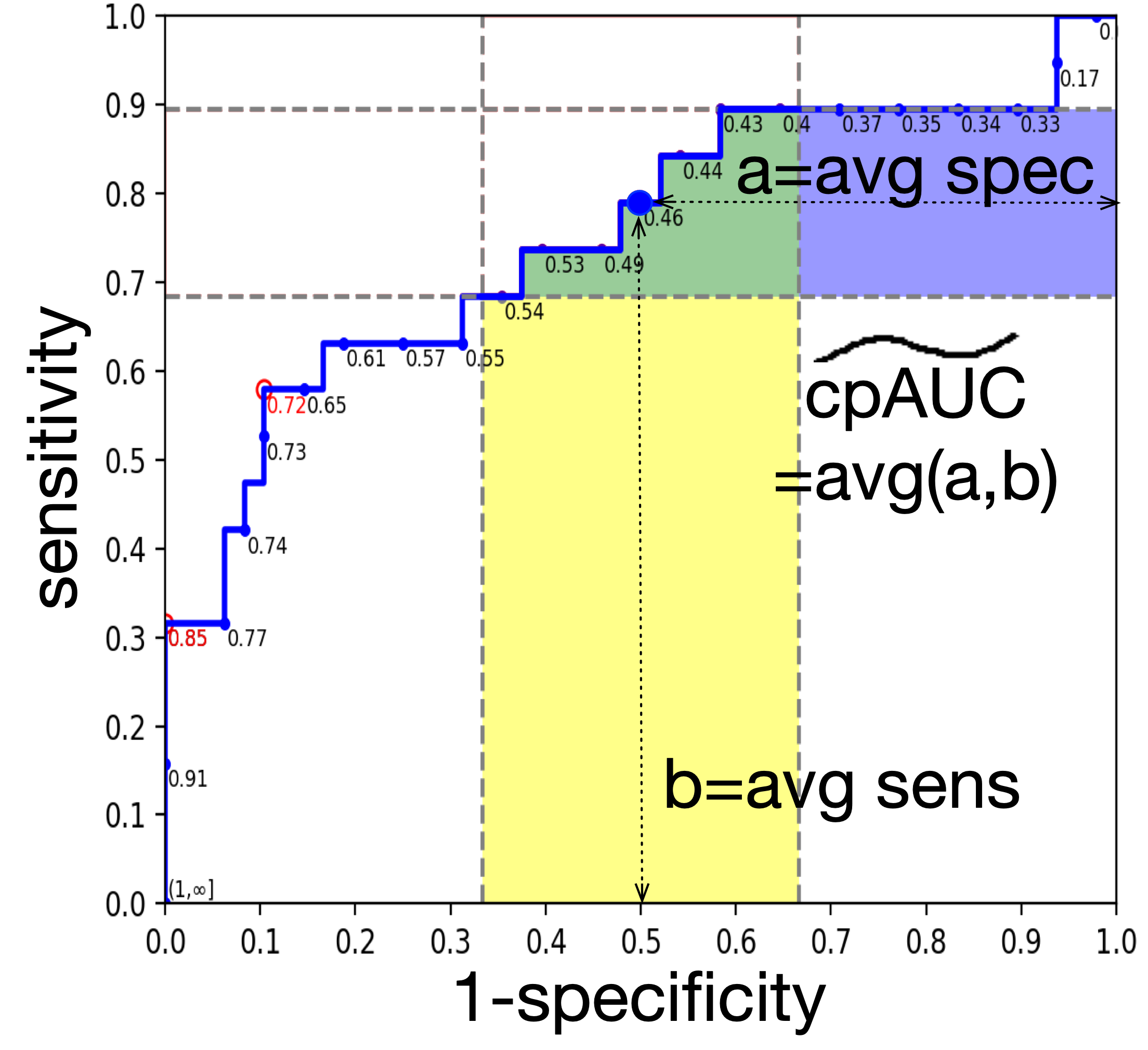}
\captionsetup{margin=10pt}
\caption{\label{fig:cpAUC_Low_Medium} The normalized concordant partial AUC $\widetilde{cpAUC}$, is illustrated for the middle third of the ROC plot. It has the same meaning and range as the $AUC$--it is a generalization thereof. It is interpreted as balanced average accuracy. It combines a vertical (yellow) and horizontal (blue) perspective.}
\end{figure}

Carrington \textit{et al.} define the concordant partial AUC ($cpAUC$) and partial C statistic ($C_\Delta$), as fully and properly analogous to the AUC and C statistic \cite{CarringtonEtAl:2020:AUC}---as generalizations, in fact. When normalized, $\widetilde{cpAUC}$ (Figure \ref{fig:cpAUC_Low_Medium}) with $\theta=(x_1,x_2,y_1,y_2)$ is interpreted as the AUC in that part, and can be compared to the AUC or $\widetilde{cpAUC}$ of any other part:
\begin{align}
     \widetilde{cpAUC}(\theta) &=
     \frac{1}{2\Delta x} \int_{x_1}^{x_2}r(x)\ dx + \frac{1}{2\Delta y} \int_{y_1}^{y_2}{1 - r^{-1}(y)}dy
     \label{eq:cpAUCn}
\end{align}
We use this measure in our method, along with a new finding: that $\widetilde{cpAUC}$ and AUC are balanced average accuracy (Sections \ref{sec:AUC-as-BAA} and \ref{sec:cpAUCn-as-BAA}). ROC curves that go above and below the major diagonal yield positive values for $\widetilde{cpAUC}$, i.e., they are properly handled.

We note that the averages in the above measures are averages of continuous values, integrals in fact, within a group of predicted risk.  They are \textbf{not} averages over multiple experiments or cross-validation folds.

\section{Method: Deep ROC Analysis}
We propose deep ROC analysis for binary classifiers, diagnostic tests, or binary prognosis at a time point to improve or confirm model selection, understanding and explanations. Our method examines measures of discrimination in greater detail, within groups.  It may complement calibration measures (if the same groups are used) and it does not include clinical utility (or rewards as utility), which may be evaluated separately. A Python toolkit\footnote{https://github.com/Big-Life-Lab/partial-AUC-C} for the method is provided for general use and limitations of the method are discussed in a later section.

\subsection{Design rationale}
We have several typical objectives when evaluating model performance:
\begin{enumerate}[label=\Alph*.]
\item To measure detection of the outcome of interest (positives); and
\item To include, rather than ignore or under-weigh, detection of the other class (negatives). Otherwise too many false positives may occur.
\item To measure pre-test and post-test detection rates.
\item To compare a model against other models.
\item To know whether the model commits more errors in detection for some groups compared to others, especially for the most relevant group(s).
\end{enumerate}
To ensure detection of the outcome, we examine sensitivity (pre-test) to understand what proportion of actual positives will be detected.  Pre-test measures are easy to understand and they are "concrete"--i.e., their direct effect on errors is obvious.

To ensure that test results are good we also need post-test measures. Positive predictive value (PPV) also called precision, is a popular \cite{ozenne2015precision,Powers2007,Altman1994ppv} and concrete measure, and it is easy to understand: it measures how often a positive test result is correct. However, PPV can be misused \cite{Worster2002}: it is only informative for low prevalence; and conversely, negative predictive value (NPV) is only informative for high prevalence. We therefore consider if likelihood ratios \cite{sokolova2009systematic,Worster2002,deeks2004lr} are better---and we conclude that it depends on the goal.

If the goal is to evaluate a test in its real-world effect with prevalence \cite{Altman1994ppv, Worster2002}, or treat patients\footnote{Worster \textit{et al.}'s states that predictive values (PPV and NPV) relate to the "probability of disease in an individual patient"}, %
or explain outcomes and errors\cite{Altman1994ppv}, then predictive values, PPV or NPV, are recommended.

If the goal is to select a test based on its intrinsic strength \cite{sokolova2009systematic,Worster2002,deeks2004lr}, without regard for prevalence and the real world effect on results and errors, possibly to account for different settings with different prevalence\cite{deeks2004lr}, then likelihood ratios are recommended.
Likelihood ratio positive (LR+) measures the detection of true positives relative to false positives, but it is not easy to understand \cite{Worster2002} because it deals with odds.

To include detection of the negative class, we use a combined pre-test measure that is (evenly) balanced in its consideration of the positive and negative class, e.g., the AUC, which is balanced average accuracy.  In part of an ROC curve, AUC is measured by the normalized concordant partial AUC, which is also balanced average accuracy. For a combined post-test measure, the diagnostic odds ratio is a logical measure associated with the likelihood ratios.

Objectives D and E, that compare performance, require a combined measure, and for that we also use AUC and the normalized concordant partial AUC, as balanced average accuracy.

Accuracy, is not a good alternative to AUC or sensitivity because, for low prevalence, accuracy obscures the outcome (inadequate for both) and for high prevalence it weighs the outcome too much (inadequate as an alternative to AUC).

\subsection{Steps in the Method}
Our method has the following steps:
\begin{enumerate}
\item Identify the purpose. To evaluate model performance:
\begin{enumerate}
    \item in general, or
    \item in general and for specific groups of patients (or instances) by predicted risk or probability
\end{enumerate}

\item Decide whether the group boundaries are by
\begin{enumerate}
    \item percentiles of FPR (or its complement, specificity), or
    \item percentiles of TPR, i.e., sensitivity, recall, or
    \item percentiles of the predicted risk or probability
\end{enumerate}

\item Decide how many groups to use and their boundary values. There should be at least 25 patients (preferably 50 or more) in each group. 

\item Create a table of average pre-test and post-test measures (e.g., Tables \ref{tab:Example-table1} and \ref{tab:Wernly-table1}) or plot measures (e.g., Figure \ref{fig:gbsg_svm_rf}). These complement an ROC plot (Figure \ref{fig:Wernly-ROC-PRC}).

\item Measure how well the model detects positives and negatives, with pre-test measures. Evaluate which models perform best and sufficiently:
\begin{enumerate}
    \item in the most relevant group(s), measured by average sensitivity, assuming the outcome is of primary interest
    
    \item in the most relevant group(s), by "AUC within the group": the concordant partial AUC (balanced average accuracy), as a combined measure, to include negatives (and avoid too many false positives)
    
    \item in a manner that is even across groups, or that gradually favours relevant group(s), measured by "AUC within the group": the concordant partial AUC (balanced average accuracy)
    
    \item overall, measured by AUC (balanced average accuracy)
\end{enumerate}

\item Measure how often a test result is correct, with post-test measures, in absolute terms with PPV (or in relative terms with LR+).   Evaluate which models perform best and sufficiently:
\begin{enumerate}
    \item in the most relevant group(s), measured by average PPV (or LR+), assuming the outcome is of primary interest and the prevalence is low. For high prevalence use NPV (or LR-).
    
    \item in the most relevant group(s), measured by balanced predictive value (or the odds ratio), as a combined measure, to include negatives.
\end{enumerate}

\item It is highly recommended to produce a calibration plot \cite{wynants2020prediction}.  If the groups within that plot align to deep ROC analysis then the plot and analysis may be compared directly.
\end{enumerate}

\subsection{Calibrated scores}
Calibrated scores for models are generally recommended. In binary classification and diagnostic testing, models not only estimate outcomes, they also output classification scores that are used to create ROC curves. Classification scores for some machine learning models are not \textbf{probabilistic} by default, yet probabilities are meaningful for interpretation. 

By default, a support vector machine produces scores in the range $[-\infty, +\infty]$ or $[a, b], a,b\in \mathbb{R}$ and some neural networks produce scores in the range $[a, b], a,b\in \mathbb{R}$. 
Calibration turns non-probabilistic scores into probabilities and improves measures of calibration \cite{Steyerberg2009b}, e.g., calibration plots and calibration in the large.

Calibration \cite{niculescu2005predicting} is an extra stage of processing that uses isotonic regression \cite{dykstra1982algorithm,mair2009isotone} or Platt's
method \cite{platt1999probabilistic,lin2007note}. It may be built into the model's implemented function or it may be available separately.

Finally, probabilistic or calibrated scores are required for option 2c in our method. Classification scores from logistic regression \cite{bishop2006pattern} and naive Bayes \cite{bishop2006pattern} are probabilistic (based on model assumptions) but they may not be well calibrated if those assumptions are not correct. Calibration can help in that case.

\section{Case Study: Mortality prediction based on arterial blood gas analysis of septic patients\label{sec:case1}}

\begin{table}[btp]
\caption{\label{tab:Wernly-table1}The neural network (abbreviated as LSTM) performs consistently well in balanced average accuracy across groups of risk by FPR: [0, 0.33], [0.33, 0.67], [0.67, 1]. Average sensitivity $\textrm{avgSens}_\theta$ is always maximal at right, while average specificity $\textrm{avgSpec}_\theta$ is always maximal at left, for equally-sized subgroups.}
\centering{}\medskip{}
\begin{tabular}{l|llll}
\hline
ROC horizontal axis (FPR):                & Global & Left          & Mid       & Right \\
                                          & [0,1]  & [0,.33]       & [.33,.67] & [.67,1] \\
Probability/risk group:                   & All    & High          & Med       & Low   \\
\hline
\textbf{LSTM}\\
Bal Avg Accuracy\,\,\, = $AUC$            & 0.88   &                &          &      \\
Group Bal Avg Acc = $\widetilde{cpAUC}$   & 0.88   & \textbf{0.89}  & 0.85     & 0.87 \\
Group Avg Sens \quad = $\widetilde{pAUC}$ & 0.88   & 0.76           & 0.91     & \textbf{0.97} \\
Group Avg Spec \quad = $\widetilde{pAUCx}$& 0.88   & \textbf{0.94}  & 0.57     & 0.20 \\
Positive predictive value                 & \multicolumn{4}{c}{0.60 at a point (t=0.5)}\\
Negative predictive value                 & \multicolumn{4}{c}{0.96 at a point (t=0.5)}\\
\hline
\end{tabular}
\end{table}

\begin{table}[btp]
\caption{\label{tab:Wernly-table2}LR performs slightly better than Lactate, but not adequately and SOFA performs poorly in groups of risk by FPR. SOFA performs best in the wrong group.}
\centering{}\medskip{}
\begin{tabular}{l|llll}
\hline
ROC horizontal axis (FPR):                 & Global & Left           & Mid       & Right \\
                                           & [0,1]  & [0,.33]        & [.33,.67] & [.67,1] \\
Probability/risk group:                    & All    & High           & Med       & Low   \\
\hline
\textbf{LR}\\
Bal Avg Accuracy\,\,\, = $AUC$             & 0.82   &                 &          &      \\
Group Bal Avg Acc = $\widetilde{cpAUC}$    & 0.82   & \textbf{0.85}   & 0.81     & 0.76 \\
Group Avg Sens \quad = $\widetilde{pAUC}$  & 0.82   & 0.67            & 0.84     & \textbf{0.94} \\
Group Avg Spec \quad = $\widetilde{pAUCx}$ & 0.82   & \textbf{0.93}   & 0.67     & 0.40 \\
Positive predictive value & \multicolumn{4}{c}{0.48 at a point (t=0.5)}\\
Negative predictive value & \multicolumn{4}{c}{0.95 at a point (t=0.5)}\\
\hline
\textbf{Lactate}\\
Bal Avg Accuracy\,\,\, = $AUC$             & 0.80   &                 &          &      \\
Group Bal Avg Acc = $\widetilde{cpAUC}$    & 0.80   & \textbf{0.81}   & 0.80     & 0.80 \\
Group Avg Sens \quad = $\widetilde{pAUC}$  & 0.80   & 0.58            & 0.88     & \textbf{0.94} \\
Group Avg Spec \quad = $\widetilde{pAUCx}$ & 0.80   &  \textbf{0.91}  & 0.65     & 0.14 \\
Positive predictive value & \multicolumn{4}{c}{-}\\
Negative predictive value & \multicolumn{4}{c}{-}\\
\hline
\textbf{SOFA}\\
Bal Avg Accuracy\,\,\, = $AUC$             & 0.72   &                 &          &      \\
Group Bal Avg Acc = $\widetilde{cpAUC}$    & 0.72   & 0.67            & 0.72     & \textbf{0.84} \\
Group Avg Sens \quad = $\widetilde{pAUC}$  & 0.72   & 0.39            & 0.82     & \textbf{0.94} \\
Group Avg Spec \quad = $\widetilde{pAUCx}$ & 0.72   & \textbf{0.80}   & 0.60     & 0.44 \\
Positive predictive value & \multicolumn{4}{c}{0.23 at a point (t=0.5)}\\
Negative predictive value & \multicolumn{4}{c}{0.92 at a point (t=0.5)}\\
\hline
\end{tabular}
\end{table}


Wernly \textit{et al.} \cite{wernly2020machine} provide a useful illustration of the need for partial area measures of discrimination in subgroups. They compare four different machine learning and clinical algorithms to predict the $32.4\%$ of septic patients who would pass away within the next $96$ hours in a multi-center ICU observational study. They evaluate a recurrent neural network using long-short term memory (LSTM) on arterial blood gas (ABG) data against several baseline models and clinical scales: Logistic Regression (LR), the SOFA score evaluating functioning of six organs, and against blood lactate levels as a sole predictor. We normalize the partial area measures reported by Wernly \textit{et al.} (Tables \ref{tab:Wernly-table1}, \ref{tab:Wernly-table2}) for interpretation.

First, we examine AUC as an overall measure. SOFA has an AUC, or average balanced accuracy of $0.72$ or $72\%$ (Table \ref{tab:Wernly-table2}), which is moderately predictive \cite{wernly2020machine}, while lactate and Logistic Regression perform well with an AUC of $80\%$ and $82\%$ respectively and LSTM’s AUC of $88\%$ (Table \ref{tab:Wernly-table1}) is $6\%$ better than others in absolute terms.

In other words, LSTM is $88\%$ accurate on average (on balance) in detecting positives and negatives without regard for class imbalance or prevalence. The AUC of $88\%$ is the average of average sensitivity at $88\%$ and average specificity $88\%$ where these measures are \textbf{necessarily equal} for the whole ROC curve, but generally different for a partial ROC curve \cite{CarringtonEtAl:2020:AUC} (Table \ref{tab:Wernly-table1}, “whole” column).

\begin{figure}[btp]
\centering
    \includegraphics[width=2.6in,height=2.5in]{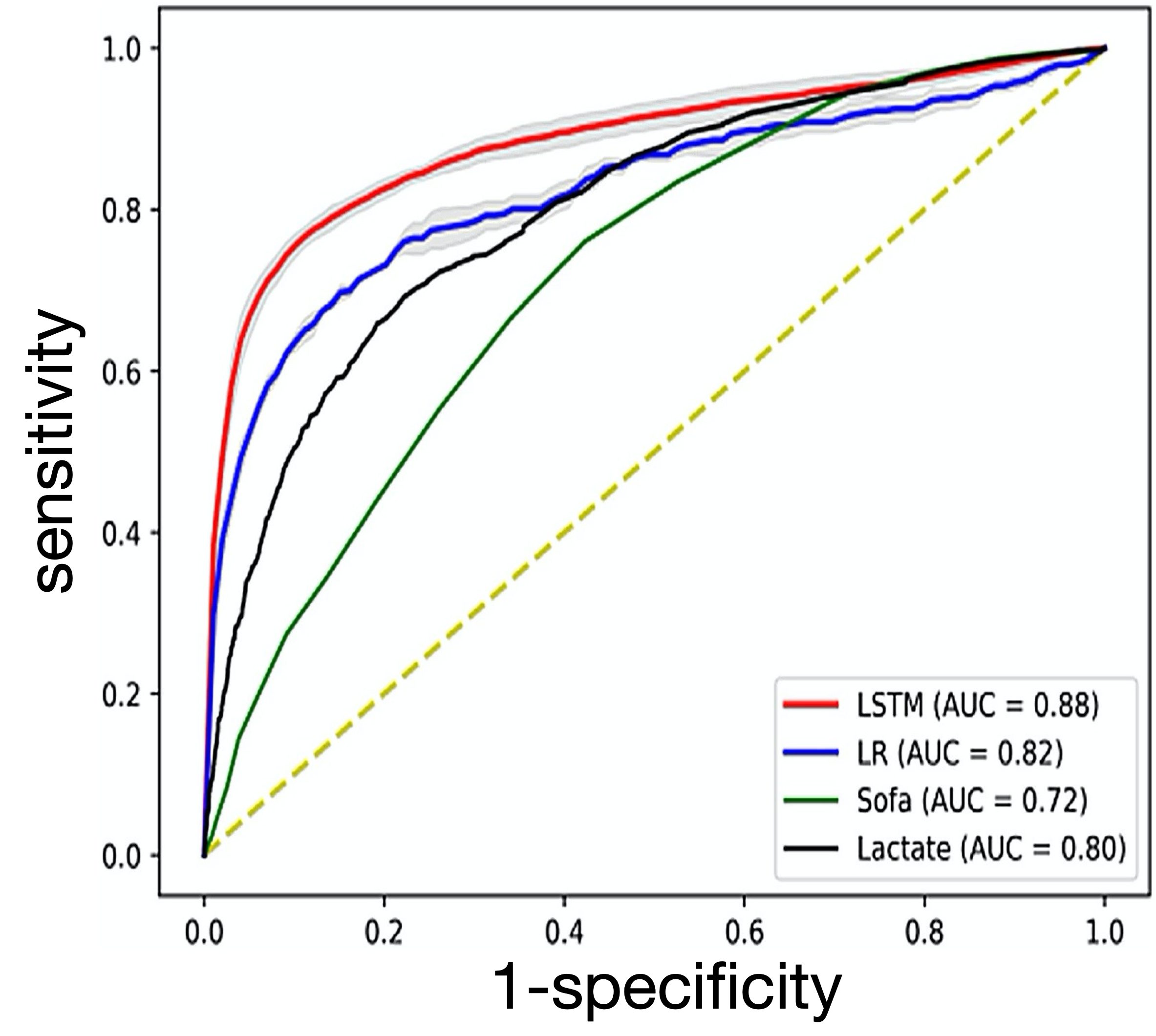}
\captionsetup{margin=10pt}
\caption{\label{fig:Wernly-ROC-PRC}The ROC plot for the four classifiers: LSTM, LR, Lactate and SOFA. LSTM is almost fully dominant.}
\end{figure}

In the ROC plot (Figure \ref{fig:Wernly-ROC-PRC}) for FPR<0.35, the curves are above each other (better) in the same order as the AUC values, but for FPR>0.5 Lactate is better than LR, and at FPR>0.63 SOFA is better than LR too.

Wernly \textit{et al.} \cite{wernly2020machine} indicate that high-risk patients are the most clinically relevant: predicting patients with “poor prognosis” and predicting with “high accuracy, with low false-positive rates”. Hence, our split of data into high, medium and low risk thirds by FPR, seems reasonable. Rather than eyeballing averages from the plot, we quantify the concordant partial AUC (or average balanced accuracy) in that region as $0.67$, $0.81$, $0.85$ and $0.89$ for SOFA, Lactate, LR and LSTM.  This means that in the region that is most relevant: SOFA is $5\%$ worse than what the AUC indicates while LR is $3\%$ better, and both LSTM and Lactate are $1\%$ better. 

\begin{figure*}[btp]
\captionsetup[subfigure]{labelformat=brace}
\subfloat[Support Vector Machine performance with a Mercer sigmoid kernel sags by about $3\%$ and $9\%$ in the high risk and low risk groups (left and right) respectively. ]{
\centering
    \includegraphics[width=2.6in,height=2.4in]{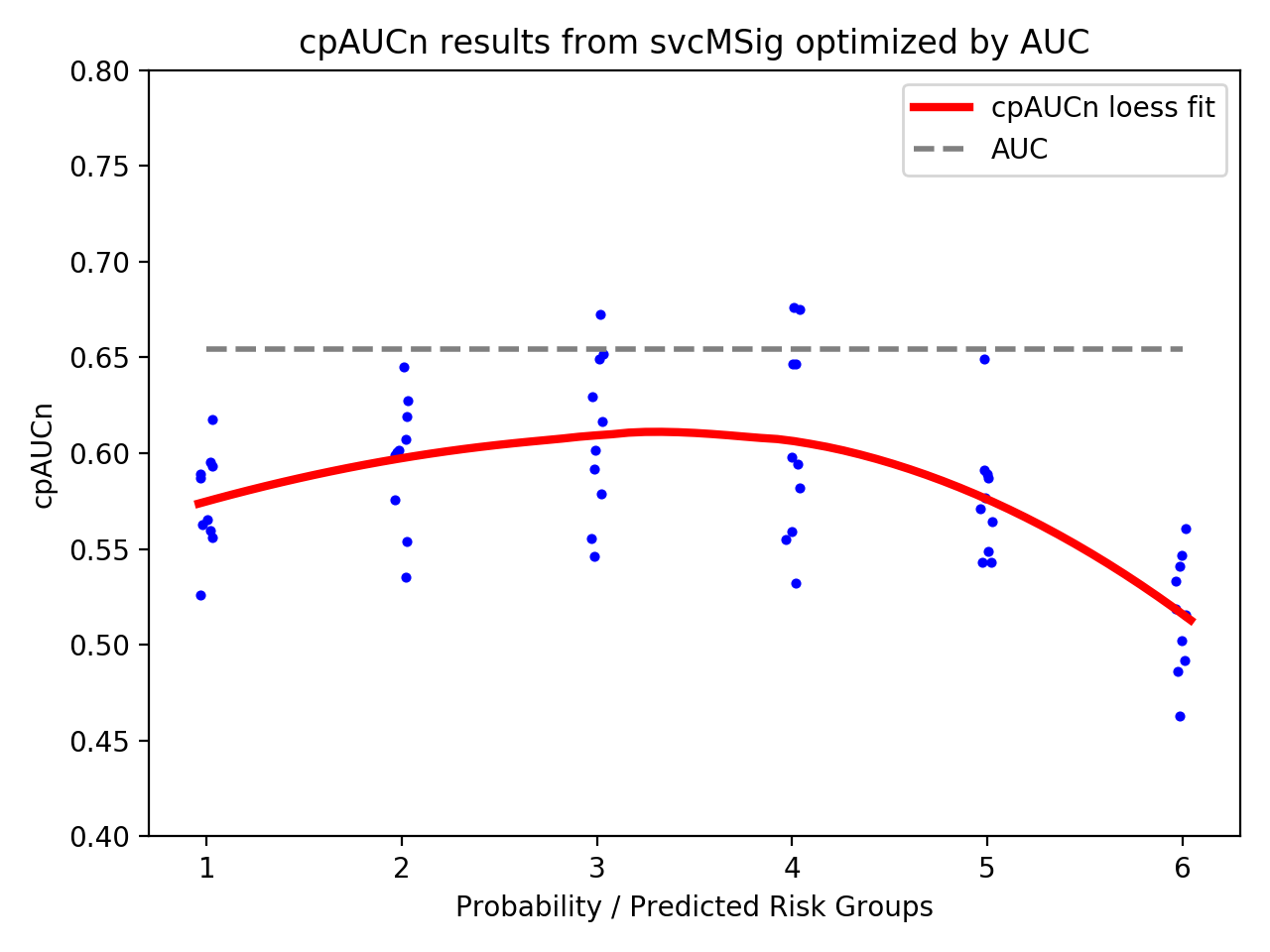}
}
\hskip2em
\captionsetup[subfigure]{labelformat=brace}
\subfloat[Random Forests performance with a small batch size sags approximately $5-6\%$ in the high and low risk groups.]{
\centering
    \includegraphics[width=2.6in,height=2.4in]{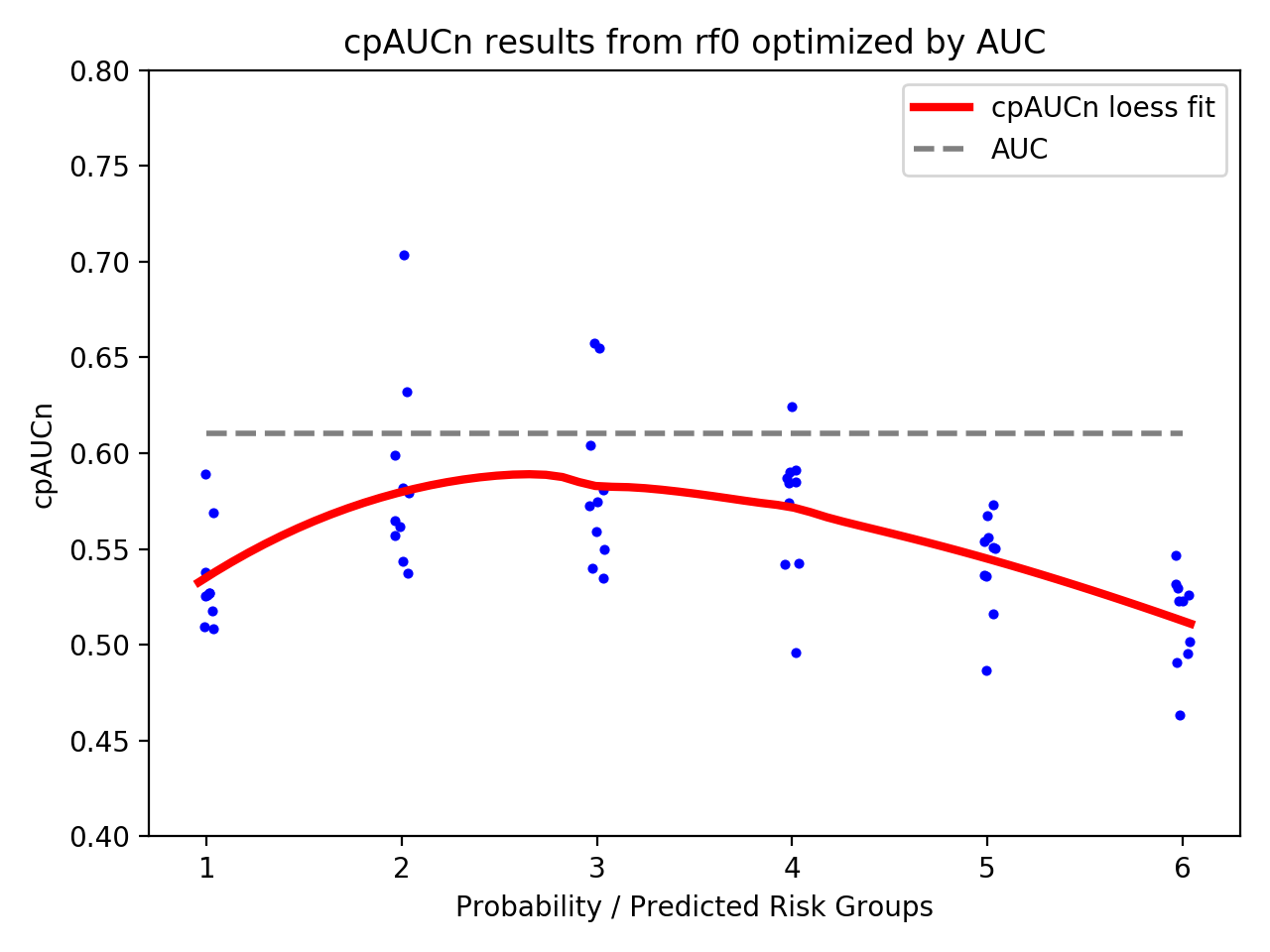}
}
\centering
\captionsetup{margin=10pt}
\caption{For the German Breast cancer Study Group data the performance of all models sags at both extents of 6 risk groups shown along the x-axis. Points from each of 10 folds are jittered for visual clarity. We explain in text the reason why the group measures (red fitted line) are below the overall measure (dashed grey line). \label{fig:gbsg_svm_rf}}
\end{figure*}

If we examine average sensitivity in the high-risk region, the differences between algorithms grow.  Between LSTM and LR, an overall difference in AUC (average balanced accuracy) of $6\%$, and a high-risk difference in the same concept ($\widetilde{cpAUC}$ = average balanced accuracy) of $4\%$, hides a $9\%$ difference in average sensitivity, which is arguably more important than average specificity and average balanced accuracy for this scenario.  That said, it is important to have the complete set of measures—the complete picture and it is helpful to report and compare $\widetilde{cpAUC}$ against the AUC value.

In absolute terms, in the high-risk region, LSTM and LR are $76\%$ and $67\%$ sensitive (on average) while AUC paints a rosier picture.  Lactate has $58\%$ average sensitivity, which is not that good, while SOFA is $39\%$ sensitive on average, which is terrible and worse than chance.

The poor sensitivity of SOFA is striking, but it makes sense. That is, in high-risk patients, there will be a lot of morbidity or organ dysfunction which SOFA identifies, e.g., if creatinine rises from 1.0 to 2.0 mg/dL.  However, a rise in creatinine from 3.0 to 6.0 mg/dL might not reflect the same importance; and the same concept applies to bilirubin, coagulation, etc. This underscores the merit of risk stratification tools with higher granularity, as in the proposed ABG-LTSM rather than SOFA. Scores such as SOFA or qSOFA or lactate concentrations were developed to “rule in” high-risk patients. However, the approach by Wernly \textit{et al.} is different, they want to “rule out” patients who are very unlikely to benefit from further critical care. SOFA performs best ($\widetilde{cpAUC}$) where it matters least (Table \ref{tab:Wernly-table2}), while Lactate performs consistently across all 3 risk groups.

LR and LSTM perform best in the high-risk region (Table \ref{tab:Wernly-table2}).  Average sensitivity and average specificity, individually cannot be compared across risk groups because they monotonically increase and decrease, respectively, from high-risk to low-risk (left to right).  

It is important to repeat that the “partial AUC” is a misnomer.  For any data, given equally sized bins, the partial AUC whether normalized ($\widetilde{pAUC}$) or not ($pAUC$) will always have the best value in the rightmost bin. If one interprets it like the AUC, then they will erroneously conclude that LSTM is most accurate overall in the rightmost (low risk) region.  The concordant partial AUC \cite{CarringtonEtAl:2020:AUC} ($\widetilde{cpAUC}$) is the proper analogue to AUC.

\section{Case study 2: German Breast Cancer Study Group\label{sec:case2}}

In survival analysis of patients in the German breast cancer study group \cite{schumacher1994rauschecker}, $33\%$ of patients with positive node primary breast cancer had isolated locoregional recurrence at 2 years after treatment. For this low prevalence situation, the minority of positives are most clinically relevant---i.e., high-risk patients identified by the leftmost part of the ROC plot.  

We applied 14 models with many different hyperparameters to predict recurrence: support vector machines (3 kernels), random forests (3 batch sizes), penalized logistic regression (4 loss functions), shallow neural networks (2 activation functions), k-nearest neighbors and decision trees.  Based on experiments we focused our analysis on three models: (i) the top performing algorithm by AUC, a support vector machine (SVM) with a Mercer sigmoid kernel \cite{carrington2014new}, (ii) a common statistical algorithm, penalized logistic regression with ridge/L1 loss, and (iii) a random forest model with a small batch size.

We computed and stored results in a 4-dimensional matrix:
\begin{itemize}
\item 100 iterations/points in hyperparameter optimization\footnote{Because of the efficiency of Bayesian search optimization, we found in experimentation that 100 iterations was sufficient--with only slight gains achieved on occasion by using 200 iterations instead. Alternative methods such as random search or grid search require more iterations.}
\item 10 folds in 2 x 5-fold cross-validation
\item 6 groups of probability or predicted risk
\item 15 group measures
\end{itemize}

We found that there was no significant difference in AUC between the best SVM model and the best random forests model when we tested the difference between matched pairs over 10 folds.  Similarly between SVM and penalized logistic regression there was no significant difference in AUC. However, in the high risk group, group 1, there was a significant difference between SVM and random forests in both cpAUCn (balanced average accuracy) and pAUCn (average sensitivity), using the same test. And the plot for logistic regression (not shown) dipped in the center and in that group it was significantly different from SVM. Hence, in some situations, deep ROC analysis leads to different decisions for model selection than than standard ROC analysis.

In the plots of performance (Figures \ref{fig:gbsg_svm_rf}a and b) there is a noticeable feature: the average of group measures are below the overall measure for all groups in this case study.  This is in contrast to the previous case study where the high risk group exceeded overall performance in two cases.

We verified that this behaviour occurs with very small and simple datasets---as was used for this case study. There were 686 samples with 228 positives split into 5 folds. When we use 6 groups for deep ROC analysis, errors in the minority class as a proportion of instances are exaggerated.  AUC is balanced average accuracy, so the exaggerated minority class errors weigh as much as majority class errors. 

We then tested whether or not using group measures for the objective in Bayesian search optimization of hyperparameters would yield better performance in absolute terms. Instead of optimizing for AUC or AUPRC, we tried optimizing $\widetilde{cpAUC}$ in group 1, and also optimizing $\widetilde{pAUC}$ in group 1.  The results (Table 4) show no significant difference in average performance in any of the four measures.

\begin{minipage}[btp]{3.2in}
\vspace{0.1in}
TABLE 4. Results for a Support Vector
Machine with a Mercer Sigmoid kernel. *No
maxima are significant.
\begin{Verbatim}[fontsize=\small]
1. Bayesian search maximizing AUC
Max mean_AUC     65.43 +/- 3.28 is at index 64 
Max mean_AUPRC   55.60 +/- 7.24 is at index 13
Max mean_cpAUCn1 58.37 +/- 2.62 is at index 34
Max mean_pAUCn1  22.85 +/- 4.51 is at index 34 *
Max mean_avgPPV  46.79 +/- 3.14 is at index 13
Max mean_avgNPV  75.53 +/- 1.50 is at index 70

2. Bayesian search maximizing AUPRC
Max mean_AUC     65.53 +/- 3.84 is at index 0
Max mean_AUPRC   55.37 +/- 7.10 is at index 51 
Max mean_cpAUCn1 58.44 +/- 2.77 is at index 86 *
Max mean_pAUCn1  22.67 +/- 4.22 is at index 24
Max mean_avgPPV  46.86 +/- 3.36 is at index 73
Max mean_avgNPV  75.68 +/- 1.64 is at index 3  *

3. Bayesian search maximizing pAUCn.group1
Max mean_AUC     65.74 +/- 3.57 is at index 29 *
Max mean_AUPRC   55.68 +/- 7.21 is at index 62 *
Max mean_cpAUCn1 58.29 +/- 2.61 is at index 67 
Max mean_pAUCn1  22.62 +/- 4.47 is at index 49 
Max mean_avgPPV  46.89 +/- 3.54 is at index 47 *
Max mean_avgNPV  75.56 +/- 1.60 is at index 87

4. Bayesian search maximizing cpAUCn.group1
Max mean_AUC     65.39 +/- 3.65 is at index 69
Max mean_AUPRC   55.17 +/- 7.00 is at index 0
Max mean_cpAUCn1 58.34 +/- 2.70 is at index 54
Max mean_pAUCn1  22.60 +/- 4.53 is at index 54
Max mean_avgPPV  46.84 +/- 3.44 is at index 15
Max mean_avgNPV  75.59 +/- 1.60 is at index 59
\end{Verbatim}
\vspace{0.7in}
\end{minipage}

Lastly, the difference in values between AUPRC and average PPV pertains to the fact that the former is weighted by change/regions in TPR (or recall as in the PRC plot) while the latter is weighted by change/regions in FPR.

\section{AUC is Balanced Average Accuracy\label{sec:AUC-as-BAA}}

We show that AUC, or AUC within a part, known as the concordant partial AUC, are interpreted as balanced average accuracy, an average of aggregate measures, which is different from average balanced accuracy (Section \ref{sec:AUC_not_ABA}), an average of point measures. First we provide definitions and notation, and then we demonstrate our claim, which is similar to a proof.

The average of a function $f(z)$ for a continuous domain $z\in\mathcal{Z}$, in the range $\theta_z=[z_1,z_2]$, is the Riemann integral, divided by the size of the range $\Delta z = z_2 - z_1$ as in (\ref{eq:continuous-averageA}).
\begin{align}
	avg_{\theta_z}\ f(z)
	&=\frac{1}{\Delta z}\int_{z_1}^{z_2}{f(z)}dz \label{eq:continuous-averageA}
\end{align}

We use x and y in the following equations to represent the axes of an ROC plot, leading to the following typical definitions for an ROC curve and AUC \cite{Dodd2003,CarringtonEtAl:2020:AUC}, from a vertical perspective:
\begin{align}
    y &= r(x) = sens(x)                \label{eq:r_xA}  \\
  AUC &= \int_{0}^{1}r(x)\ dx          \label{eq:AUCyA} \\
      &= \int_{0}^{1}sens(x)\ dx       \label{eq:AUCyA2}
\end{align}

AUC (\ref{eq:AUCyA},\ref{eq:AUCyA2}) equals average sensitivity \cite{ZhouEtAl:2002:StatisticalBook, CarringtonEtAl:2020:AUC}---i.e., (\ref{eq:AUCyA}) is in the form of (\ref{eq:continuous-averageA}) with a normalization factor $\nicefrac{1}{\Delta z}=1$.

An ROC curve and AUC are also defined as follows \cite{Dodd2003,CarringtonEtAl:2020:AUC}, from the horizontal perspective:
\begin{align}
    x &= r^{-1}(y) = 1 - spec(y)       \label{eq:r_yA}  \\  
  AUC &= \int_{0}^{1}{1 - r^{-1}(y)}dy \label{eq:AUCxA} \\
      &= \int_{0}^{1}{spec(y)}dy       \label{eq:AUCxA2}
\end{align}

AUC (\ref{eq:AUCxA}) equals average specificity \cite{Jiang1996, ZhouEtAl:2002:StatisticalBook, CarringtonEtAl:2020:AUC}.

It follows from (\ref{eq:AUCyA}) and (\ref{eq:AUCxA}) that AUC must equal the average of the two (for $\Delta x, \Delta y=1$):
\begin{align}
  AUC &= \frac{1}{2} \int_{0}^{1}r(x)dx
	    +\frac{1}{2} \int_{0}^{1}1 - r^{-1}(y)dy
	     \label{eq:AUCxy} \\
      &= \frac{1}{2}\int_{0}^{1}sens(x)dx 
	    +\frac{1}{2}\int_{0}^{1}spec(y)dy
	     \label{eq:AUCxy2} \\
	  &= avg \left[\ avg_{\Delta x}(sens(x)) + avg_{\Delta y}(spec(y)) \ \right]
	     \label{eq:AUCxy3}
\end{align}

We call the above, balanced average accuracy, because it is the balance (average) of average accuracy in each class. We further justify this interpretation as follows.

For AUC, which refers to a whole ROC curve, any weighted average of average sensitivity and average specificity is equal to AUC, because the two parts are equal, but only the simple average generalizes to a partial ROC curve \cite{CarringtonEtAl:2020:AUC}, discussed in the next section. 

We can see the similarity in form between (\ref{eq:AUCxy2}), and balanced accuracy (\ref{eq:balanced_accuracy_w}), $b$, at a point $w$:
\begin{align}
	b(w) &= \frac{1}{2} sens(w) 
	       +\frac{1}{2} spec(w)	 \label{eq:balanced_accuracy_w} \\
	     &= avg \left[\ sens(w) + spec(w) \ \right]
	     \label{eq:balanced_accuracy_w2}
\end{align}
Equations (\ref{eq:AUCxy3}) and (\ref{eq:balanced_accuracy_w}) have similar interpretations---the former, AUC, refers to \textbf{average} sensitivity and \textbf{average} specificity over the whole ROC curve, while the latter, balanced accuracy, refers to sensitivity and specificity at a point. Hence the name and interpretation. 

We further note that Youden's index $J$ at a point $w$ is related to $b(w)$:
\begin{align}
    J(w) = 2b(w) - 1 \label{eq:Youden}
\end{align}

\section{The Normalized Concordant Partial AUC is Balanced Average Accuracy\label{sec:cpAUCn-as-BAA}}
The previous concepts also apply to part of an ROC curve, or one of multiple groups of risk.  In such a part or group, AUC is called the normalized concordant partial AUC $\widetilde{cpAUC}$, and we show that it is also Balanced Average Accuracy. $\widetilde{cpAUC}$ is defined as follows \cite{CarringtonEtAl:2020:AUC} for $\Delta x=x_2-x_1$ and $\Delta y=y_2-y_1$:
\begin{align}
  \widetilde{cpAUC} &= \frac{1}{2 \Delta x} \int_{x_1}^{x_2}r(x)dx
   	      +\frac{1}{2 \Delta y} \int_{y_1}^{y_2}1 - r^{-1}(y)dy
	       \label{eq:cpAUC} \\
        &= \frac{1}{2 \Delta x}\int_{x_1}^{x_2}sens(x)dx 
	      +\frac{1}{2 \Delta y}\int_{y_1}^{y_2}spec(y)dy
	       \label{eq:cpAUC2} \\
	  &= avg \left[\ avg_{\Delta x}(sens(x)) + avg_{\Delta y}(spec(y)) \ \right]
	     \label{eq:cpAUC3}    
\end{align}
We can again see the similarity in form between (\ref{eq:cpAUC2}), (\ref{eq:balanced_accuracy_w}) and (\ref{eq:AUCxy2}).  It therefore yields the same interpretation: $\widetilde{cpAUC}$ is Balanced Average Accuracy as the balance (or average) of \textbf{average} sensitivity and \textbf{average} specificity for part of an ROC curve.

\section{\label{sec:AUC_not_ABA}AUC is not Average Balanced Accuracy}
We have shown that AUC is the balance of average accuracy for each class: the average of average sensitivity and average specificity, computed as an integral (or computed discretely at each point\footnote{We do not prove the discrete form because it is complex and meticulous, but it is easily seen in every experimental result using our deepROC Python toolkit}).  However, AUC is not the average of balanced accuracy at each point. We show this with a simple example (Figs. \ref{fig:AUC-not-ABA1}).

\begin{figure}[btp]
\centering{
\includegraphics[scale=0.48]{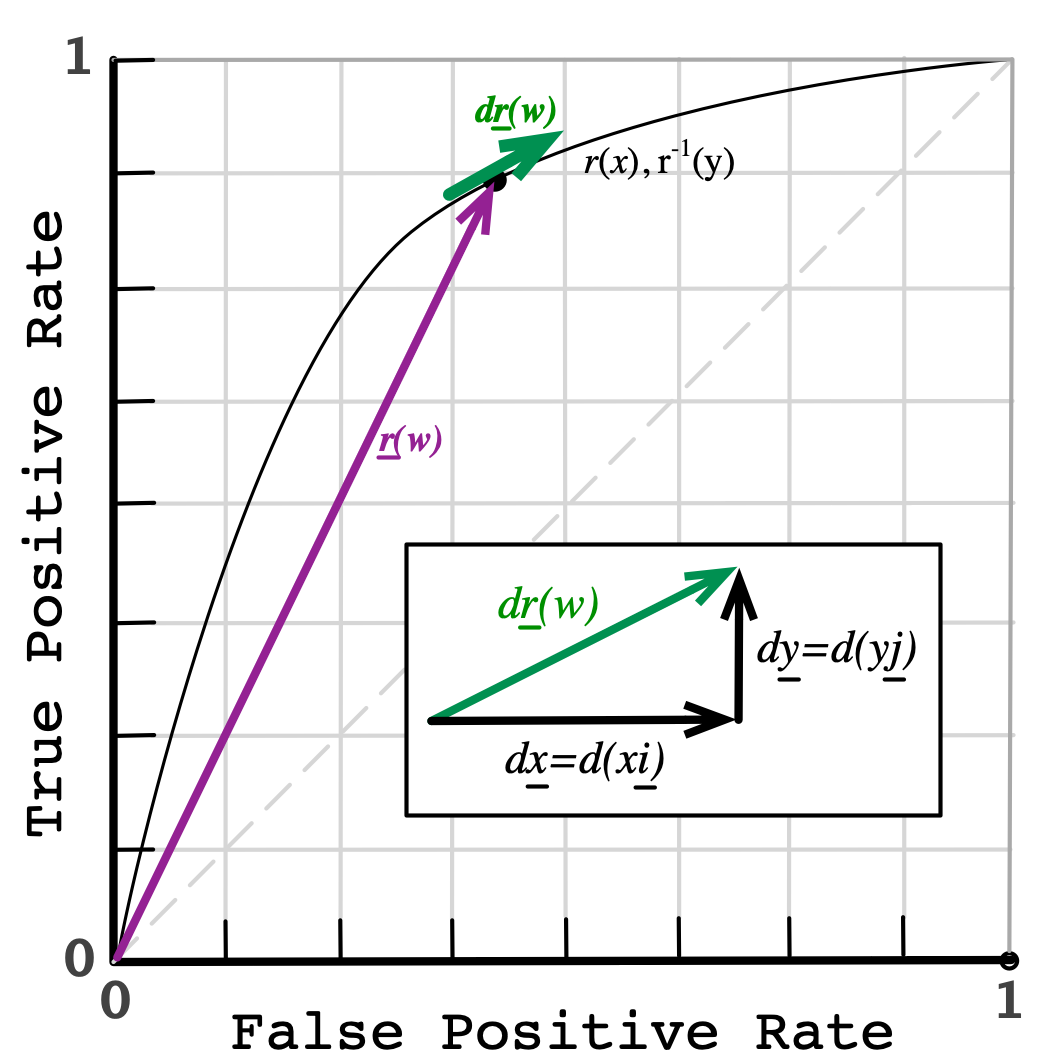}}
\caption{The vector aspect of balanced accuracy at points $w$ along an ROC curve.}
\label{fig:b_complexity}
\end{figure}

\begin{figure*}[btp]
\centering{
\includegraphics[scale=0.48]{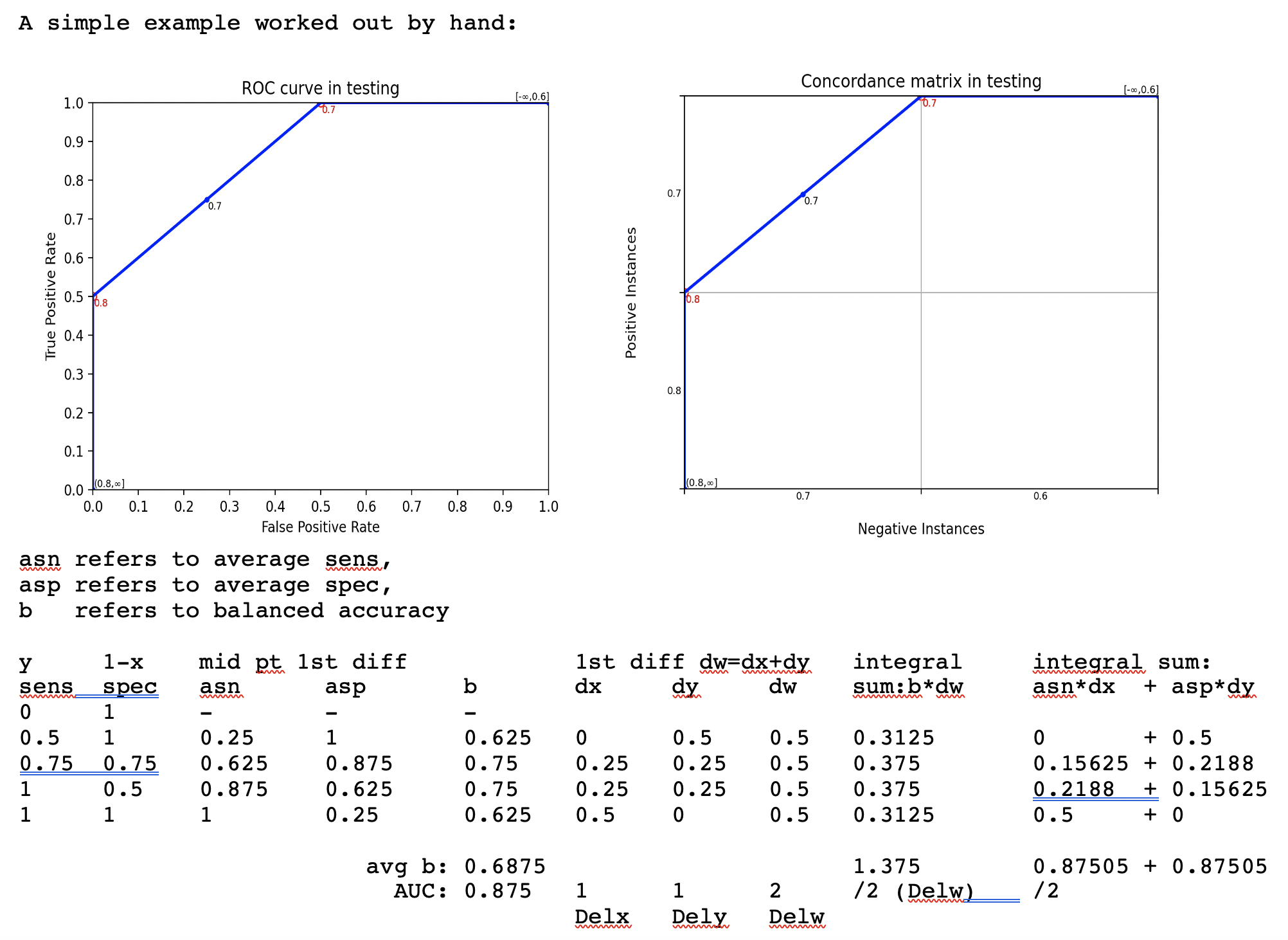}}
\caption{A simple example of AUC as balanced average accuracy not average "balanced accuracy".}
\label{fig:AUC-not-ABA1}
\end{figure*}

It may also help to understand the difference in terms of equations as follows. From balanced accuracy (\ref{eq:balanced_accuracy_w}) and a continuous average (\ref{eq:continuous-averageA}) we can express average balanced accuracy for a range $\theta_{w}=[w_1,w_2]$ where $w$ is a continuous index along the ROC curve:
\begin{align}
	avg_{\theta_w}\ b(w)
	&=\frac{1}{\Delta w}\int_{w_1}^{w_2} {b(w)}\ dw \notag \\
    &=\frac{1}{\Delta w}\int_{w_1}^{w_2}{
      \frac{1}{2} (sens(w) + spec(w)) }\ dw       \label{eq:avg-b-wholeA}
\end{align}
Figure \ref{fig:b_complexity} shows the vector nature of $dw$ in (\ref{eq:avg-b-wholeA}):
\begin{align}
    \Delta w &= \Delta sens(w) + \Delta spec(w) \notag \\
             &= |\Delta y| + |\Delta x|
             \label{eq:Delta_w}\\
          dw &= |dy| + |dx| \notag \\
             &= dy - dx 
             \label{eq:dw}\\
           k &= \frac{1}{2 \Delta w}
\end{align}

This leads to complexity instead of expressions equal to AUC, such as the following:
\begin{align}
	avg_{\theta_w}\ b(x,y;w)  
	=k \int_{w_1}^{w_2}{ sens(w) }\, [dy - dx] \notag \\
    +k \int_{w_1}^{w_2}{ spec(w) }\, [dy - dx] \notag \\
    =k \int_{y_1}^{y_2}{ sens(y) }\, dy
    -k \int_{x_1}^{x_2}{ sens(x) }\, dx \notag \\
    +k \int_{y_1}^{y_2}{ spec(y) }\, dy
    -k \int_{x_1}^{x_2}{ spec(x) }\, dx             
\end{align}

Hence, AUC is not average balanced accuracy, but is balanced average accuracy.

\section{Limitations\label{sec:limitations}}
One possible limitation of our method is that the additional information introduces more complexity which could complicate communication of results. Providing guidance to ensure uniform reporting is recommended, wherever possible. 

For small datasets that do not meet the preferred threshold of 50 instances or more per group, as in our second case study, we observed that errors bias group performance downward compared to overall measures of performance.

Another limitation is that our method pertains to binary classifiers and diagnostic tests, including prognosis at a time point with binary outcomes. Since some authors deride dichotomization, even when and where appropriate for decision-making, we discuss this point in further detail in the following section.
\section{Discussion on Continuous Methods versus Predicted Risk and Subgroups in Binary Classification}

An opinion piece by Wynants \textit{et al.} \cite{Wynants:2019:Myths} argues against thinking or methods that bin, categorize or group data with continuous values. It is opinion because philosophically it is up to the clinician to decide whether or not binning or categorizing helps them make decisions or not. Some of the present authors posit that many detection, diagnosis, prognosis or treatment decision-making problems are categorical in nature (e.g., the patient has strep throat) while other problems in the same categories may be ordinal or continuous (e.g., what dosage to apply). 

If one is confident that clinicians and people are diligent and capable of keeping both a category and a number in mind, then there is no concern. Ideally we would all be free to choose our own tools and judge or present evidence in our own way.

It has been argued \cite{Wynants:2019:Myths} that categorization loses information, and that is true in terms of information entropy, but not all information is useful. If there is too much unnecessary information, then the signal-to-noise ratio is limited and our understanding and decision-making suffers---i.e., summarizing and categorizing is useful. Summary descriptions are the essence of the word "statistic".  Hence, binary classification and subgroups of predicted risk have a role to play.

That said, there are limits to binary and categorical thinking in clinical prediction models--they assume a set of options and tests known a priori, i.e., completeness.  However, diagnosing and formulating a therapy for a patient may not be well-defined (explicit) nor complete. Differential diagnosis and treatment, or other decisions, may go beyond any medical protocols and order sets (if/when they exist).  Decisions may or may not fall within routine experience and treatment.  Diagnosis and decision-making may involve generating, synthesizing and investigating treatment options not previously considered by the clinician; and the dynamic nature of decision-making may involve a clinician's gut feel based on continuous values using Bayesian thinking.
In this context, some are concerned that binary classification and categorical/subgroup methods might distract or blind a clinician, regulator, etc.

\section{Conclusions and Future Work}
We have shown that models (or tests) can and do behave differently in different groups of risk---and within those groups their performance that may be better or worse than the average overall. Our method may identify needs for applications using AUC where no deficiencies have been perceived \cite{Borji:2021:Saliency},\cite{WangLing:2021:Salient}.    

We have demonstrated that the normalized concordant partial AUC ($\widetilde{cpAUC}$) as balanced average accuracy is useful to interpret a model's performance in each group--it indicates where an algorithm is strong or weak. Our new interpretation of AUC is also helpful since it applies to individuals in contrast to the pairwise interpretation of AUC as a C statistic. 

In the first case study, LSTM model aside: our method more clearly differentiates LR versus Lactate in the high risk group that matters most, and it more clearly shows the inadequacy of SOFA in absolute terms. Hence, deep ROC analysis can improve model selection in some cases and it provides an informed view of model performance by groups for assurance.  

In the second case study, we observed how the model performs differently in groups with lesser performance in the highest and lowest risk groups for that data.  We used Bayesian search optimization with group measures ($\widetilde{cpAUC_n}$ and $\widetilde{pAUC_n}$) as objectives instead of AUC or AUPRC but there was no significant difference in results.  Future work could examine if group measures would improve optimization for large datasets.


\section*{List of abbreviations}
\begin{tabular}{ll}
AI:             & Artificial intelligence \\
\textit{AUC}:   & Area under the ROC curve \\
\textit{AUPRC}: & Area under the precision recall curve \\
$C$:            & The $C$ statistic for binary outcomes,\\ 
                & but not Harrell or Uno's $C$ statistic \\
$C_\Delta$:     & The partial $C$ statistic \\
\textit{FNR}:   & False negative rate \\
\textit{FPR}:   & False positive rate, or 1-specificity \\
$pAUC$:         & Partial area under the ROC curve (i.e., vertical)\\
$\widetilde{pAUC}$: & Normalized partial area under the ROC curve \\
$cpAUC$:        & Concordant partial area under the ROC curve \\
$\widetilde{cpAUC}$:& Normalized concordant partial area under the \\
                & ROC curve \\
$pAUCx$:        & Horizontal partial area under the curve \\
$\widetilde{pAUCx}$:& Normalized horizontal partial area under the\\
                & ROC curve \\
LR:             & Logistic Regression \\
LSTM:           & Long Short-Term Memory \\
PAI:            & Partial area index \\
PRC:            & Precision recall curve \\
ROC:            & Receiver operating characteristic \\
SOFA:           & Sequential organ failure assessment \\
$sPA$:          & Standardized partial area \\
\textit{TNR}:  & True negative rate, or specificity, or selectivity \\
\textit{TPR}:  & True positive rate, or sensitivity, or recall \\
xAI:           & Explainable artificial intelligence \\
\end{tabular}

\section*{Availability of code and data}
The Python code that produced the measurement numbers, plots and tables, is available at:
\begin{verbatim}https://github.com/Big-Life-Lab/deepROC\end{verbatim}
\begin{verbatim}http://deepROC.org\end{verbatim}

\noindent The German Breast Cancer data is available at:
\begin{verbatim}https://biostat.app.vumc.org/wiki/Main/DataSets\end{verbatim}



\section*{Acknowledgements}
Parts of this work has received funding by the Austrian Science Fund (FWF), Project: P-32554 ``A reference model for explainable Artificial Intelligence in the medical domain''.


\ifCLASSOPTIONcaptionsoff
\fi



%

{
\bibliographystyle{IEEEtran}
\bibliography{references}
}


%

\begin{IEEEbiography}[{\includegraphics[width=1in,height=1.25in,clip,keepaspectratio]{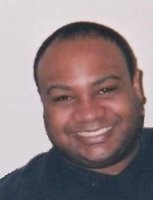}}]{Andr\'e M. Carrington} is a Post Doctoral Research Fellow at the Ottawa Health Research Institute. Andr\'e received his Ph.D. in Systems Design Engineering and Masters in Mathematics (Computer Science) from the University of Waterloo. 
\end{IEEEbiography}

\vspace{-1cm}

\begin{IEEEbiography}[{\includegraphics[width=1in,height=1.25in,clip,keepaspectratio]{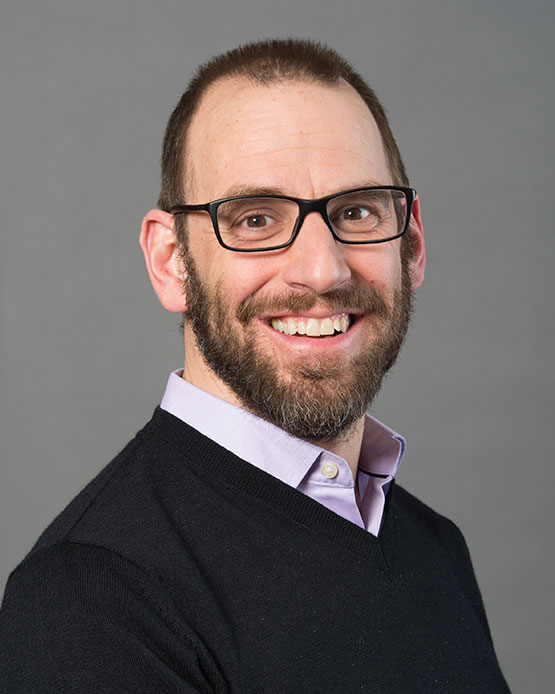}}]{Douglas G. Manuel} is a Medical Doctor with a Masters in Epidemiology and Royal College specialization in Public Health and Preventive Medicine. He is a Clinician Scientist at the Ottawa Hospital Research Institute and the Bruyère Research Institute and a Professor in the Departments of Family Medicine and the School of Epidemiology, Public Health and Preventive Medicine at the University of Ottawa.
\end{IEEEbiography}

\vspace*{-1cm}

\begin{IEEEbiography}[{\includegraphics[width=1in,height=1.25in,clip,keepaspectratio]{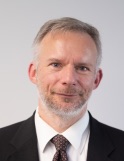}}]{Paul W. Fieguth} is a Professor in Systems Design Engineering and Associate Dean in the Faculty of Engineering at the University of Waterloo and co-Director of the Vision \& Image Processing group.  Paul received his Ph.D.\ in electrical engineering from the Massachusetts Institute of Technology. 
\end{IEEEbiography}

\vspace*{-1cm}

\begin{IEEEbiography}[{\includegraphics[width=1in,height=1.25in,clip,keepaspectratio]{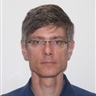}}]{Tim Ramsay} is the head of the Ottawa Methods Centre at the Ottawa Hospital Research Institute and a Professor with the University of Ottawa, Canada. 
\end{IEEEbiography}

\vspace*{-1cm}

\begin{IEEEbiography}[{\includegraphics[width=1in,height=1.25in,clip,keepaspectratio]{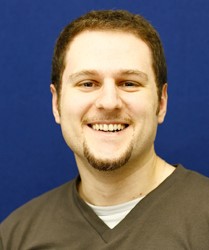}}]{Venet Osmani} is a Senior Researcher in the eHealth Group at the Fondazione Bruno Kessler Research Institute and a Professor in the Department of Psychology and Cognitive Science, University of Trento, Italy.
\end{IEEEbiography}

\vspace*{-1cm}

\begin{IEEEbiography}[{\includegraphics[width=1in,height=1.25in,clip,keepaspectratio]{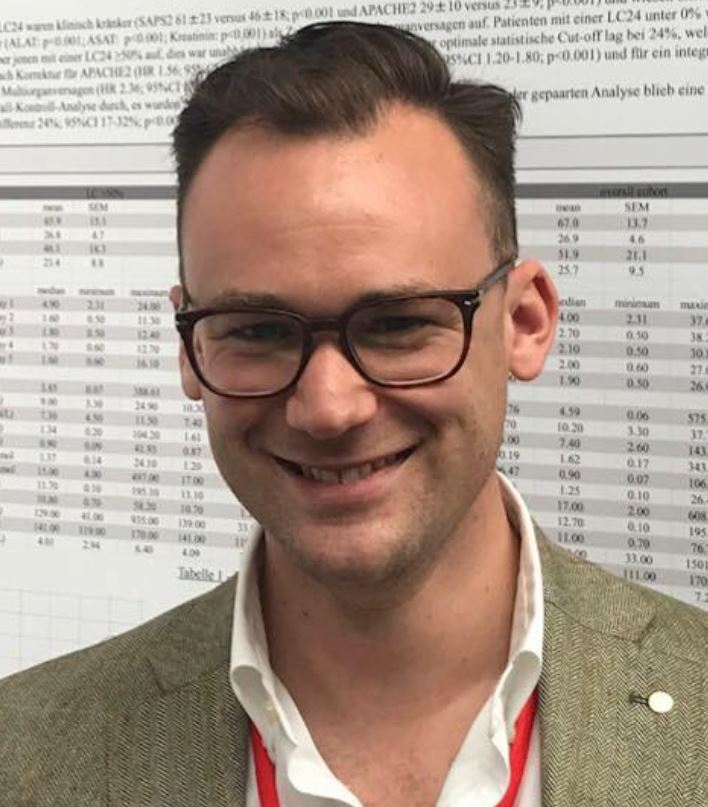}}]{Bernhard Wernly} practices internal medicine in the Department of Cardiology at the Paracelsus Medical University of Salzburg, Salzburg, Austria.
\end{IEEEbiography}

\vspace*{-1cm}

\begin{IEEEbiography}[{\includegraphics[width=1in,height=1.25in,clip,keepaspectratio]{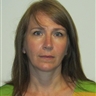}}]{Carol Bennett} is a research associate at the Ottawa Hospital Research Institute and the Institute for Clinical Evaluative Sciences, Ottawa, Canada. 
\end{IEEEbiography}

\vspace*{-1cm}

\begin{IEEEbiography}[{\includegraphics[width=1in,height=1.25in,clip,keepaspectratio]{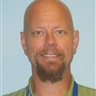}}]{Steve Hawken} is the head of Big Data initiatives in the Ottawa Methods Centre at the Ottawa Hospital Research Institute and a Professor with the University of Ottawa, Canada. 
\end{IEEEbiography}

\vspace*{-1cm}

\begin{IEEEbiography}[{\includegraphics[width=1in,height=1.25in,clip,keepaspectratio]{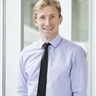}}]{Matt McInnes} is a radiologist with the Ottawa Hospital Research Institute and a Professor with the University of Ottawa, Canada. 
\end{IEEEbiography}

\vspace*{-1cm}

\begin{IEEEbiography}[{\includegraphics[width=1in,height=1.25in,clip,keepaspectratio]{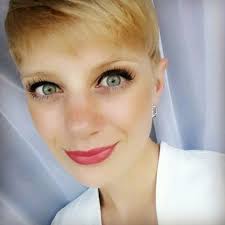}}]{Olivia Magwood} is a research associate at Bruyère Research Institute and a doctoral student at the University of Ottawa. She has a Master's degree in public health.
\end{IEEEbiography}

\vspace*{-1cm}

\begin{IEEEbiography}[{\includegraphics[width=1in,height=1.25in,clip,keepaspectratio]{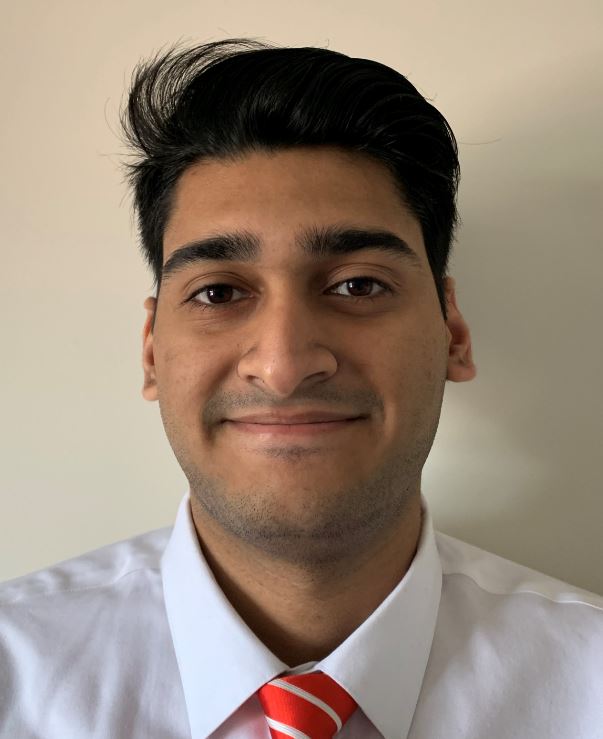}}]{Yusuf Sheikh} is a part-time researcher at the Ottawa Hospital Research Institute. Yusuf is completing his Bachelor's degree in Biomedical Science at the University of Ottawa, Canada. 
\end{IEEEbiography}

\vspace*{-1cm}

\begin{IEEEbiography}[{\includegraphics[width=1in,height=1.25in,clip,keepaspectratio]{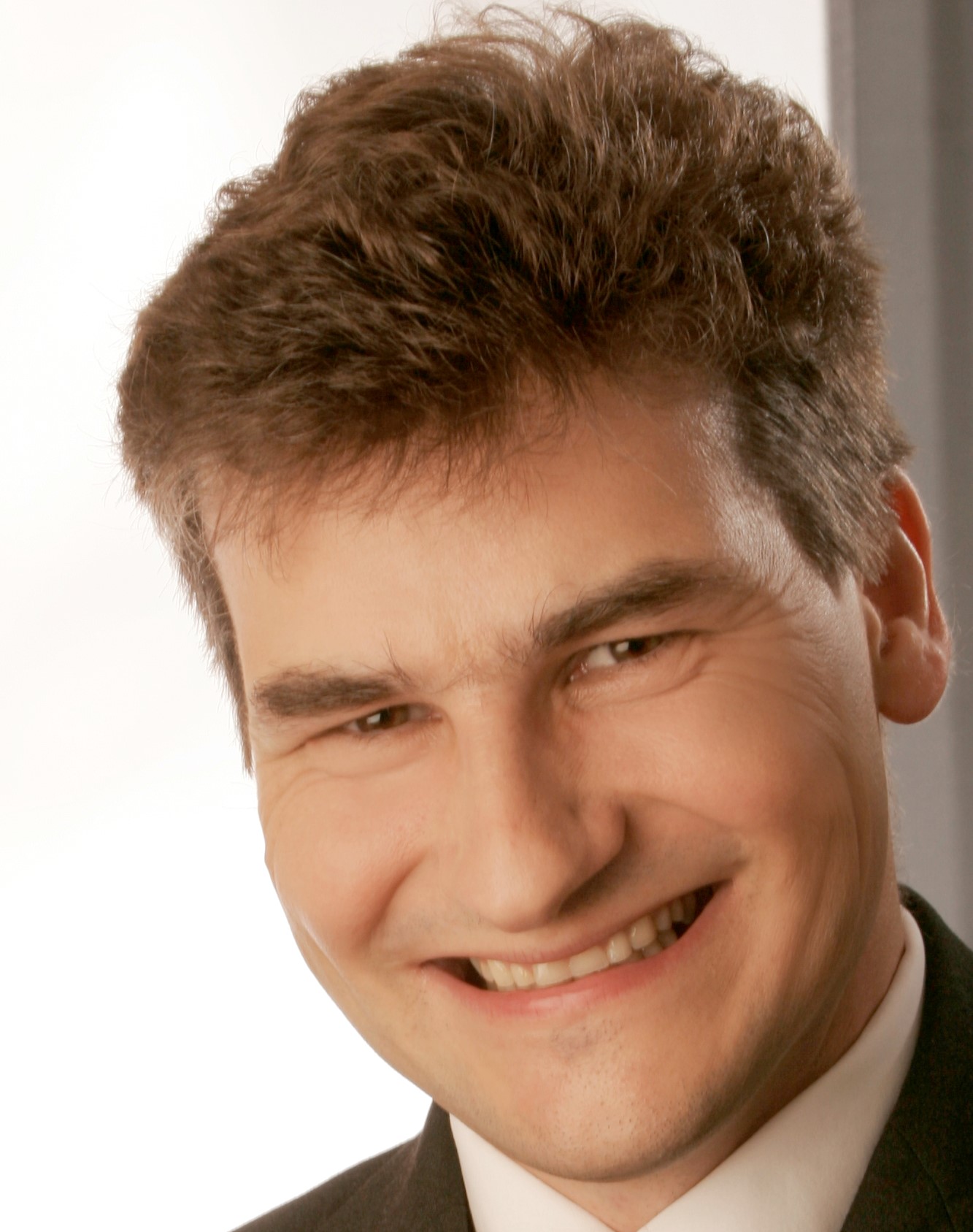}}]{Andreas Holzinger} (M'00) is Visiting Professor for explainable AI at the University of Alberta, Canada since 2019 and head of the Human-Centered AI Lab at the Medical University Graz, Austria. He received his PhD in cognitive science from Graz University and his second PhD in computer science from Graz University of Technology. Andreas promotes a synergistic approach to put the human-in-control of AI to align it with human values, privacy, security and safety.
\end{IEEEbiography}

\end{document}